\documentclass[acmsmall]{acmart}

\setcopyright{cc}
\setcctype{by}
\acmDOI{10.1145/3832110}
\acmYear{2026}
\acmJournal{PACMSE}
\acmVolume{3}
\acmNumber{ISSTA}
\acmArticle{ISSTA019}
\acmMonth{10}
\acmSubmissionID{issta26main-p172-p}
\received{2026-01-30}
\received[accepted]{2026-06-25}

\usepackage{newtxtext,newtxmath}
\usepackage{amsmath,amsfonts}

\usepackage{hyperref}
\usepackage{subcaption}
\newsavebox{\mybox}

\newcommand{\revisioncolor}{black}
\newcommand{\revision}[1]{{\color{\revisioncolor}#1}}

\newenvironment{revisionblock}
  {\begingroup\color{\revisioncolor}}
  {\endgroup}

\newcommand{\toolname}{\textsc{Remi}\xspace}
\newcommand{\toolnameLimi}{\textsc{LIMI}}
\newcommand{\toolnameExpGA}{\textsc{ExpGA}\xspace}
\newcommand{\toolThemis}{\textsc{Themis}\xspace}
\newcommand{\toolAFT}{\textsc{AFT}\xspace}

\newcommand{\toolLimiWCite}{\toolnameLimi~\cite{LIMI-XiaoISSTA2023-10.1145/3597926.3598099}}
\newcommand{\toolExpGAWCite}{\toolnameExpGA~\cite{fanExpGAICSE22}\xspace}
\newcommand{\toolThemisWCite}{\toolThemis~\cite{angell2018themis}\xspace}
\newcommand{\toolAFTWCite}{\toolAFT~\cite{AFT-zhao-ase2024}\xspace}

\DeclareMathOperator{\E}{\mathbb{E}}

\usepackage{multirow}
\usepackage{listings}
\usepackage[shortlabels]{enumitem}

\usepackage[ruled,vlined,linesnumbered,lined]{algorithm2e}
\usepackage{algpseudocode}
\usepackage{graphicx}

\newsavebox{\answerboxbox}
\newenvironment{answerbox}{%
  \begin{lrbox}{\answerboxbox}%
    \begin{minipage}{0.97\linewidth}%
    \ignorespaces
}{%
    \end{minipage}%
  \end{lrbox}%
  \noindent\fcolorbox{gray!50}{gray!5}{\usebox{\answerboxbox}}%
}

\newsavebox{\grayframebox}

\usepackage{tabu}
\usepackage{wrapfig}

\begin{document}

\title{Fairness Invariants: A Relational Approach to Explaining and Mitigating Fairness Bugs}

\author{Ranit Debnath Akash}
\orcid{0009-0004-6694-9670}
\affiliation{%
  \institution{University of Illinois at Chicago}
  \city{Chicago}
  \country{USA}
}
\email{rakas@uic.edu}

\author{Ashish Kumar}
\orcid{0000-0001-8773-2084}
\affiliation{%
  \institution{Pennsylvania State University}
  \city{University Park}
  \country{USA}
}
\email{azk640@psu.edu}

\author{Gang Tan}
\orcid{0000-0001-6109-6091}
\affiliation{%
  \institution{Pennsylvania State University}
  \city{University Park}
  \country{USA}
}
\email{gtan@psu.edu}

\author{Saeid Tizpaz-Niari}
\orcid{0000-0002-1375-3154}
\affiliation{%
  \institution{University of Illinois at Chicago}
  \city{Chicago}
  \country{USA}
}
\email{saeid@uic.edu}

\begin{abstract}
Data-driven software systems are increasingly deployed in high-stakes socio-economic domains, from criminal justice to financial lending. However, these systems often exhibit individual discrimination---unjustified disparities in which a program yields different outcomes for similar individuals who differ only in their protected attributes (e.g., race, gender, age). While existing research has focused on detecting and quantifying these bugs, there remains a critical lack of principled mechanisms to explain and localize individual fairness bugs. Current explanation techniques are largely designed for single-input decisions rather than the relational nature of discrimination, which inherently involves a comparison between an original and a counterfactual pair. 

We present \toolname, a framework for the automated localization, explanation, and mitigation of individual discrimination. Inspired by loop-invariant synthesis in formal methods, we treat counterfactual fairness as a relational invariant discovery problem. We introduce a bidirectional relational explanation framework that learns over paired examples $(x, x')$ to identify regions of the input space where fairness is violated. Unlike traditional one-way implication pairs used in invariant inference, our approach enforces bidirectional constraints: requiring identical outcomes for both original and counterfactual samples. \toolname utilizes three data-alignment techniques to infer interpretable rule-based models that act as "fairness invariants." These rules serve as guardrails to selectively block or relabel unfair predictions without requiring model retraining. Our evaluation on symbolic and neural network programs demonstrates that \toolname localizes ground-truth fairness bugs in over 83\% of cases, significantly outperforming state-of-the-art baselines and reducing discriminatory decisions in black-box models by up to 70\%.
\end{abstract}




\maketitle

\section{Introduction}
\label{sec:intro}

Automated decision-support software systems have become foundational to modern socio-economic infrastructure. 
They have been used to make critical decisions in domains such as criminal justice, healthcare, financial lending, and hiring. 
However, because these models often learn from historical datasets, they risk encoding and amplifying biases related to protected attributes like race, gender, or disability status. Hence, ensuring their fairness has emerged as a critical requirement.

\noindent In high-stakes scenarios, this manifests as "fairness bugs"---unjustified disparities where the software yields different outcomes for individuals who are identical in all relevant qualifications, but differ in a protected characteristic. For instance, studies on FICO scoring found that black non-defaulters were often assigned higher risk scores than their white counterparts~\cite{10.5555/3157382.3157469}.

\noindent In response to these risks, the software engineering and machine learning communities have developed a diverse array of techniques to detect, explain, and mitigate bias through pre-processing data~\cite{chakraborty2020fairway,chakraborty2019software}, in-processing algorithmic adjustments~\cite{tizpaz2022fairness-10.1145/3510003.3510202,gohar2023understanding,nguyen2023fix}, and post-processing calibration~\cite{10.5555/3157382.3157469}.
While these efforts have significantly improved our ability to address fairness issues, a principled mechanism for explaining individual fairness bugs remains largely lacking. Current explanation techniques primarily focus on local interpretability, explaining why a model made a specific decision for a single input~\cite{ribeiro2016should}. 
However, individual fairness is inherently relational: it cannot be violated by a single input, but requires an analysis between a specific individual $x$ and their "similar" counterpart $x'$ who may differ only in their protected attributes. We identify two core limitations. 

\begin{enumerate}
\item  \textbf{The Localization Insufficiency}: Existing tools provide explanations for model decisions (e.g., "why was this loan denied?"), but they lack the framework to explain discriminatory outcomes (e.g., "why was this loan denied to $x$, but granted to $x'$?"). We currently lack a way to consider multiple similar inputs simultaneously to pinpoint the exact logic driving the disparity.

\item \textbf{The Mitigation Problem.} Developers require compact and precise characterizations of how the counterfactual unfairness occurs. Without a way to localize these "fairness bugs" to specific relational constraints, mitigation strategies are limited and ineffective.    
\end{enumerate}

\noindent \textbf{Intuition.} Inspired by loop-invariant synthesis from the programming language literature~\cite{solar2006combinatorial}, we develop a relational explanation and mitigation technique for software fairness. 
In loop-invariant synthesis, the goal is to identify a region of the program state space that captures all reachable program states. 
The key is to enforce 'implication pairs': if a state satisfies the head of an implication pair, then it must also satisfy the tail (a positive instance). Decision tree inference with implication pairs~\cite{Garg2016dtinvct} formalizes this by classifying states into positive and negative examples. The learned decision tree must therefore respect these unidirectional relational constraints in addition to separating positive and negative instances. The relational fairness problem can be viewed as an analogous task. 

\vspace{0.25 em}
\noindent \textbf{Key Observation.} Instead of reasoning about reachable program states, we infer fairness invariants of automated decision-making software and identify regions of the input space where the original $x$ (head) and counterfactual $x'$ (tail) pairs disagree. 
Each relational pair $(x, x')$ acts as a constraint, requiring that the program assign the same outcome to both original and counterfactual samples that may only differ in their protected attributes. 
Unlike the one-directional implication pairs used in invariant inference, these counterfactual constraints are \emph{bidirectional}: if one element of the pair receives a favorable outcome, the other must as well for fairness to hold, and likewise for unfavorable outcomes. Hence, the approach for using decision trees to respect unidirectional relational constraints can be extended to bidirectional constraints, enabling us to use decision trees to solve the counterfactual fairness problem.

\vspace{0.25 em}
\noindent \textbf{Approach.} We present \toolname (Relational Explanation and MItigation), a framework to localize, explain, and mitigate individual fairness bugs. \toolname first generates counterfactual instances to identify discriminatory pairs. It then constructs a relational dataset where pairs are labeled '+' if the model treats them identically (fair) and '-' if the outcomes differ (unfair). We propose three alignment techniques---original, horizontal, and vertical extension---to structure this relational data. From this, \toolname infers interpretable rules that distinguish fair regions from unfair ones 
Finally, \toolname extracts these discriminatory rules as guardrails to selectively block or relabel unfair predictions, mitigating bias without requiring model retraining.

\vspace{0.25 em}
\noindent \textbf{Experiments.}
We perform experiments both on symbolic (rule-based) and neural network (black-box) programs.
Since the ground truth for the symbolic programs is known and verifiable, we first use \toolname to answer four research questions about i) the performance of three relational dataset curation techniques; ii) the performance of nine different rule inference algorithms; iii) the characteristics and precision of extracted rules vs. the ground truth; and iv) the performance of guardrails technique for a mitigation, based on the extracted discriminatory rules. 
We find that horizontal extension outperforms the baseline and other alignment methods, and C4.5Tree, FIGS, and CART tree inferences outperform other rule-based inference methods. The alignment and inference techniques localize the ground truth in more than 83\% of the cases, compared to the state-of-the-art AFT~\cite{AFT-zhao-ase2024} approach. Additionally, the rule-based guardrails reduce individual instances of discrimination in all cases. Finally, we study whether our experiences with \toolname over symbolic programs generalize to black-box neural networks. We found that applying guardrail rules to neural networks reduces individual discriminatory decisions by at least 40\%, up to 70\%,  significantly outperforming the retraining-based bias mitigation techniques~\cite{AFT-zhao-ase2024,fanExpGAICSE22,zhang2020white,udeshi2018automated}


\vspace{0.5em}\noindent \textit{Contributions.} The key contributions of this paper are: 
\begin{itemize}[leftmargin=*]
\item Inspired by the loop invariant synthesis, we design a data-driven approach to infer fairness invariants for automated decision-making programs,

\item We develop a novel guardrail strategy to mitigate individual discrimination,

\item We put forward \toolname, a framework that automatically detects, localizes, explains, and mitigates individual fairness bugs. 

\item Our experiments on both rule-based symbolic and data-driven neural network programs show that \toolname significantly outperformed the baseline techniques in terms of localizing and mitigating individual discrimination.

\end{itemize}

\section{Overview}\label{sec:overview}

Following our intuition that individual fairness is a relational property that cannot be violated over a single point $x$. But it requires a relational analysis between the original and counterfactual inputs ($x,x'$) that only differ in their protected attributes. In light of this individual fairness~\cite{dwork2012fairness}, a decision by decision-making programs-under-test (DPuTs) is considered unfair (discriminatory) if only flipping the protected attributes (and any logically dependent non-protected attributes) changes the outcome of DPuTs. 
We demonstrate that explaining the DPuT’s decision for a single data point, e.g., local explanation methods~\cite{ribeiro2016should}
is insufficient. 

%
\begin{figure*}
    \centering
    \includegraphics[width=1.0\textwidth]{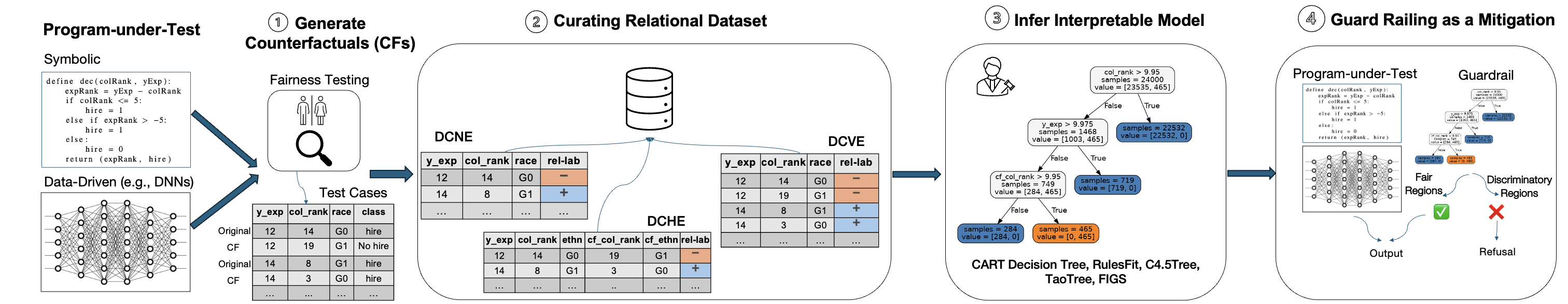}    
    \caption{\toolname Framework.}
    \label{fig:framework}
    \vspace{-1.0 em}
\end{figure*}

\noindent \textbf{\toolname Workflow Summary.} 
Here we will give an overview of how \toolname works. 
\toolname is a tool that identifies and explains individual fairness violations within decision-making programs. It identifies the specific logic that causes a program to produce discriminatory outcomes based on protected attributes, such as race. 
Our workflow consists of four main stages, as illustrated in Figure~\ref{fig:framework}. 
It has two elements: (1) the decision-making program under test (DPuT) and (2) a dataset containing protected and non-protected attributes. 

\noindent \textbf{A. Counterfactual generation and finding discriminatory instances}. The first stage is to build a relational data point. In doing so, \toolname generates a counterfactual instance for any given input by changing the protected attributes. A relational data point becomes discriminatory (unfair) when the DPuT gives different outcomes for each input in the pair.

\noindent \textbf{B. Relational dataset construction and curation.} 
Evaluating the DPuT on both $x$ and $x'$ yields a relational label: $+$ if both outcomes agree, and $-$ if they disagree (a fairness bug). 
\toolname  introduces three data curation schemes, which all keep the relational label $\mathcal{L}$ "fair ($+$) vs unfair ($-$)", but transform the features differently:

\begin{itemize}
    \item \textit{Data curation with no extension (DCNE).} This approach keeps only the original point ($x$), but the label of the point is the corresponding relational label $\mathcal{L}$ as derived by comparing the outcome of DPuT on $x$ against the counterfactual point $x'$. 
    \item \textit{Data Curation with vertical extension (DCVE).} It augments the original data point $x$ with its counterfactual ($x'$) as a separate instance. The labels of both instances are common and equal to the relational label $\mathcal{L}$. This doubles the number of training instances and exposes the tree learner to more examples of fair and discriminatory profiles. 
    \item \textit{Data Curation with horizontal extension (DCHE).} This method concatenates original ($x$) and counterfactual ($x'$) features into a single extended feature vector, keeping only the features that are not identical in both ($x_i \ne x'_i$). 
This explicitly exposes interactions between original vs counterfactual values that correlate with unfairness.
\end{itemize}

Across all the preprocessing techniques, the target is the relational label $\mathcal{L}(x,x')$, i.e., ($+$ or $-$). 
The non-protected attributes (and, for DCHE, their counterfactual counterparts) are the only inputs to the interpretable learner, so that the learned rules describe where the DPuT is unfair.

\noindent\textbf{C. Interpretable learning and rule extraction.} Given the relational dataset, \toolname instantiates the explainer $\E$ using a set of tree-based algorithms (e.g., CART, C4.5, TaoTree, RulesFit, FIGS). 
Our target is to retrieve a performant, interpretable learner that can explain what features distinguish fair relational samples from the unfair ones. 
The tree partitions the curated feature space into regions whose leaves are labeled with relational labels. 
From leaves with a high proportion of unfair relational labels, \toolname extracts if–then rules that characterize discriminatory subspaces (e.g., “age $\le$ 20.5 $\land$ priors\_count $\le$ 3.5” or “college rank = 50 $\land$ years-of-experience $\le$ 1 year”). 

\noindent
\textbf{D. Guardrail-based mitigation.} Finally, \toolname uses the extracted rules as guardrails at the final debiased model during deployment time. 
When a new input falls into an unsafe region, the framework can deny automatic prediction and defer to human review.
Because guardrails are enforced selectively on localized regions, \toolname avoids heavy-handed global post-processing that might degrade accuracy or introduce new fairness violations elsewhere.



\noindent\begin{minipage}{.45\textwidth}
\begin{lstlisting}[
language=Python,
caption={FairSquare \cite{fairsquare17oopsla} populaiton generation},
label={lst:FairsquarePopulation}
]{Name}
def popModel()
  ethnicity = gauss(0,10)
  colRank = gauss(25,10)
  yExp = gauss(10,5)
  
  if ethnicity > 10:
    colRank = colRank + 5
    
  return (colRank, yExp)
\end{lstlisting}
\end{minipage}\hfill
\begin{minipage}{.45\textwidth}
\begin{lstlisting}[language=Python, caption={FairSquare \cite{fairsquare17oopsla} hiring decision},
label={lst:FairsquareHiring}
]{Name}
def dec(colRank, yExp):
    expRank = yExp - colRank
    if colRank <= 5:
        hire = 1
    else if expRank > -5:
        hire = 1
    else:
        hire = 0
    return (expRank, hire)
\end{lstlisting}
\end{minipage}

\noindent \textbf{Overview Example}: We use an example of a hiring program, taken from Albarghouthi et al.~\cite{fairsquare17oopsla}, to illustrate our approach (see  Listings~\ref{lst:FairsquarePopulation},\ref{lst:FairsquareHiring}), where the ethnicity is a protected attribute, and the college rank of an individual is causally and directly affected by their ethnicity (e.g., being Hispanic and attending a Hispanic-Serving Institution).
Specifically, a low college rank and more years of prior job experience are key to hiring. 
The program's decision function appears fair, as it uses only college rank and years of experience to decide hiring, but an upstream generative model causally influences college rank. 

Our approach starts by generating the population (following Listing~\ref{lst:FairsquarePopulation}) and passing those samples to obtain the score and hiring decision from \texttt{dec} program~\ref{lst:FairsquareHiring}.
\toolname first generates counterfactual applicants by changing ethnicity and adjusting affected attributes, then labels pairs where the hiring decision flips as unfair.
For this example, \toolname generates 30,000 applicant samples from the program and their corresponding counterfactual with a different ethnicity. 
Our sampling finds 2,250 individual discriminatory instances, applicant pairs that differ only in ethnicity, but receive different hiring decisions (labeled `-'). Note that the rest (27,750 samples) is fair w.r.t. ethnicity and labeled `+'.
We are interested in explaining and mitigating the individual discriminatory bugs in this program.

\noindent\textbf{Which curation scheme best enables discrimination localization?}
Among the three relational dataset curation, we found that 
DCNE infers models with very high predictive metrics (accuracy around 0.99, Precision close to 0.99), leading to the localization of bugs in up to 2,234 out of the 2,250 \revision{individual discriminatory instances} (IDIs).
This convinces us that the relational labels in the training data are a key driver of root-cause localization quality.

\noindent\textbf{Which interpretable learner is the best at identifying discriminatory regions?}
We compare multiple interpretable learners (e.g., CART, C4.5, TaoTree, rule-based baselines), while fixing the data curation method to DCNE. We found that a C4.5Tree~\cite{C4.5-tree-Quinlan} achieves superior performance and localizes 2,237 IDIs among the 7 tree-based and interpretable algorithms. 

\noindent\textbf{How does our approach perform to pinpoint the root cause of the fairness bug?}
Our relational decision trees yield compact rules that localize unfair regions of interest.
In total, it retrieves 97 rules and covers 98\% of the considered discriminatory instances.
For example, one rule captures 301 of the 2250 IDIs (13.4\% discrimination coverage) is the following: \texttt{\{ 14.6<col-rank $\le$ 17.25 $\land$ 11.99 < y-exp $\le$ 14.84 $\land$ cf-ethnicity > 10.5 $\land$ 17.85< cf-col-rank $\le$ 22.95 \}} where it finds a narrow region of college rank and year of experience when the DPuT of PG1 becomes significantly discriminatory based on the ethnicity.
Intuitively, the rule recovers the root cause of the discrimination: in a narrow mid-college rank, mid-experience range 
(14.6< col-rank $\le$ 17.25, 11.99 < y-exp $\le$ 14.84), switching to an unfavorable ethnicity increases the counterfactual college rank 
(17.85 < cf-col-rank $\le$ 22.95) and flips the hiring decision. 
Because this rule explicitly captures the ethnicity-to-college rank-to-hire pathway, it demonstrates a very close match with the ground truth level root cause of the discrimination in this hiring program.

\noindent\textbf{How to Mitigate Unfairness:} 
We treat the learned rules as guardrails around the original program. On this program, inserting discriminatory rules as guardrails reduces the number of discriminatory instances from 2,250 to just 2, while preserving other similar classification metrics. We note that naively replacing the program with an unconstrained decision tree trained on counterfactual labels actually increases the number of fairness bugs to 2,430 on unseen samples.


\section{Relational Explanation and Mitigation Problem}
\label{sec:problem}
We consider any Decision-Making Program under Test (DPuT), which is essentially a function that maps an input describing an individual into a binary decision, such as \emph{favorable} (e.g., loan approved) or \emph{unfavorable} (e.g., loan denied). The input features for each individual can be grouped into two disjoint categories:
    \textit{Protected attributes} ($A_p$) such as race or gender, which by fairness considerations should not affect the decision, and
    \textit{Non-protected attributes} ($A_{np}$) such as income, age, or education level.
Formally, the DPuT can be represented as a function
\[
    f(A_p, A_{np}) \;\to\; Y, \quad Y \in \{\text{favorable}, \text{unfavorable}\}.
\]

\noindent \textbf{Fairness Notion.} We use an individual fairness definition that requires two similar individuals should receive similar outcomes irrespective of their protected attributes ~\cite{dwork2012fairness}. Following the counterfactual fairness definition~\cite{10.5555/3294996.3295162}, a decision by DPuT is considered unfair (discriminatory) for an individual if only changing the protected attributes (and any logically dependent non-protected attributes) flips the decision, i.e., $f(x_p,x_{np}) \neq f(x'_p,x'_{np})$ where $x_p \neq x'_p \land x_{np} \sim x'_{np}$.

\vspace{0.5 em}
\noindent \textbf{Relational Labeling for Discrimination.}
We are given a DPuT together with an input dataset $\mathcal{D}$. To evaluate fairness, we transform $\mathcal{D}$ into a relational dataset, where pairs of datapoints are explicitly labeled as fair or unfair. We formalize this notion below:

\begin{definition}[Relational Dataset]
A relational dataset is a pair $\mathcal{D}_R = (\mathcal{D}', R)$
where $\mathcal{D}' = \{x_1, \ldots, x_n\}$ is a set of datapoints and 
$R : \mathcal{D}' \times \mathcal{D}' \to \{+, -\}$ 
is a labeling function that assigns to each pair $(x_i, x_j)$ either $+$ (indicating a fair pair) or $-$ (indicating an unfair pair).
\end{definition}

For each datapoint $x_i = (a_{p_i}, a_{np_i}) \in \mathcal{D}$, where $a_{p_i} \in A_p$ are the protected attributes and $a_{np_i} \in A_{np}$ are the non-protected attributes, we construct a corresponding counterfactual datapoint $x'_i$. The counterfactual is obtained by modifying the protected attributes, denoted $a'_{p_i}$, while keeping the non-protected attributes unchanged, except in cases where a change in the protected attribute necessitates an adjustment to some non-protected attributes (e.g., changing sex from male to female requires adjusting the relationship attribute from ``husband'' to ``wife''). Formally,
\[
    x'_i = (a'_{p_i}, \widehat{a_{np_i}}), \quad \text{where } a'_{p_i} \neq a_{p_i},
\]
and $\widehat{a_{np_i}}$ denotes the non-protected attributes of $x_i$, possibly updated to maintain consistency with the change in $a'_{p_i}$.

We then evaluate the DPuT on both the original datapoint and its counterfactual, obtaining outcomes $y_i = f(x_i)$ and $y'_i = f(x'_i)$, where $f$ is the function modelling the DPuT. This comparison induces a relational label on the pair $(x_i, x'_i)$:
\begin{itemize}
    \item \textbf{Agreement ($+$):} The outcomes agree, indicating fair treatment.
    \[
        L(x_i, x'_i) = + \quad \text{if } y_i = y'_i
    \]
    \item \textbf{Disagreement ($-$):} The outcomes differ, indicating potential counterfactual unfairness.
    \[
        L(x_i, x'_i) = - \quad \text{if } y_i \neq y'_i
    \]
\end{itemize}
In this construction, the set $\mathcal{D}'$ of the relatioanl dataset consists of all original datapoints together with their counterfactual counterparts, and the relation $R$ is defined by the agreement or disagreement labels assigned to each pair.
The key challenge is not to explain the overall decision function $f(x)$ where $x \in \mathcal{D}$. Instead, our focus is on explaining the relational label $L(x, x')$, which captures whether a pair of datapoints is treated fairly or unfairly.

Our goal is to learn a human-interpretable model $E$ that, 
can characterize the conditions under which a pair receives a disagreement label ($-$), i.e., when the Decision-Making Program under Test (DPuT) exhibits unfair behavior.
In other words, we want to answer the question: 
\textbf{Informal Problem Statement.} \emph{For which regions of the input space, defined solely by non-protected attributes, does the DPuT exhibit counterfactual unfairness?}

\vspace{0.25 em}
\noindent \textbf{Formal Problem Statement}
Let our relational dataset be 
$$\mathcal{D}_{\text{R}} = \{ ((x_1, x'_1), L_1), ((x_2, x'_2), L_2), \dots, ((x_n, x'_n), L_n) \}$$
where each $x_i = (a_{p_i}, a_{np_i})$, $x'_i = (a'_{p_i}, \widehat{a_{np_i}})$, and the label $L_i \in \{+, -\}$ indicates whether the outcomes of the DPuT on $(x_i, x'_i)$ agree or disagree. For convenience, we denote $Z_i = (x_i, x'_i)$.

\noindent \textit{Explanation.} Our goal is to learn an interpretable explanation model $E$, instantiated as a decision-tree based learner (e.g., CART, C4.5, FIGS, RulesFit, or similar variants). The model $E$ predicts whether a pair $Z_i$ will receive an agreement ($+$) or disagreement ($-$) label. In practice, decision tree training is framed as a binary classification problem, where the learning algorithm recursively partitions the feature space so as to minimize an impurity measure at each split. For classification, the impurity at a node is most commonly measured using the \emph{cross-entropy} between the two possible classes - agreement and disagreement. If $p_{+}$ and $p_{-}$ denote the empirical probabilities of the labels $+$ and $-$, respectively, among the datapoints at a node, then the cross-entropy is then given by
\[
H(p_{+}, p_{-}) = - p_{+} \log p_{+} - p_{-} \log p_{-} 
\] 
Thus, the explanation model $E$ is a set of human-interpretable rules that delineate regions of the feature space $Z$ where the DPuT is most likely to exhibit counterfactual unfairness.

\noindent \textit{Mitigation.}
Since the learned model $E$ characterizes the discriminatory regions purely in terms of attributes, the rules describing its disagreement ($-$) regions can be
extracted as \emph{guardrails}. Let $\mathcal{R}^{-}$ denote this rule set, inducing a
guardrail predicate $g(a_{np}) = 1$ iff $a_{np}$ satisfies some rule $r \in \mathcal{R}^{-}$,
i.e., the input lies in a region where $f$ is likely to discriminate. We then deploy a
\emph{guarded} decision function $f_g(x) = f(x)$ when $g(a_{np}) = 0$, and
$f_g(x) = \bot$ when $g(a_{np}) = 1$, where $\bot$ denotes withholding the automatic
outcome and deferring the case to human review (or, alternatively, relabeling it to
restore a consistent outcome across the pair $(x, x')$). The mitigation objective is to
choose $\mathcal{R}^{-}$ so that $f_g$ minimizes the residual counterfactual
unfairness while keeping both the benign inputs it unnecessarily blocks and the loss in predictive utility small. 

\section{Approach}\label{sec:approach}

We define a region as fair if the DPuT’s outcome remains invariant under changes to protected attributes; otherwise, the region is unfair, indicating the existence of Individual Discriminatory Instances (IDIs). To localize these bugs, our tool \toolname, employs an interpretable decision tree to classify fair versus unfair regions. This model captures feature interactions and enables the extraction of precise decision rules that characterize the discriminatory input space.
As illustrated in Figure~\ref{fig:framework} and Algorithm~\ref{alg:unflocDT}, our framework consists of four primary steps:

\begin{itemize}
    \item Section~\ref{cf-gen-idi}: Generating counterfactuals to detect IDIs.
    \item Section~\ref{cur-rel-data}: Curating relational datasets from identified pairs.
    \item Section~\ref{syn-expl}: Training an interpretable model to extract explanation rules for discrimination.
    \item Section~\ref{mit-gaurd}: Deploying these rules as guardrails to mitigate unfairness.
\end{itemize}

\subsection{Counterfactual Generation and Finding Discriminatory Instances}~\label{cf-gen-idi} 
For each of our different relational dataset construction and training approaches and programs (DPuTs), this is a common step.
We take as input a decision-making program under test $\texttt{DPuT}$, which has access to sensitive attributes from the original dataset ($X_{orig}$), the program will be applied to. 
After applying the program $\texttt{DPuT}$ on $X_{orig}$, we can retrieve their target (favorable or unfavorable) outcome $Y_{orig}$ (Line 1 in Algorithm \ref{alg:unflocDT}).
Then we create a counterfactual $x'_i$ for each data point $x_i \in X_{orig}$ based on its protected attribute.
For our symbolic programs, we just flip the protected attribute (and update accordingly related dependent non-protected attribute) of  to get the corresponding counterfactual entry ($x'_i \in X_{cf}$), and retrieve the target outcome for counterfactual ($Y_{cf}$) based on the decision of DPuT (Line 2-5 in Algorithm \ref{alg:unflocDT}). 
Here $x'_i$ is the corresponding counterfactual $\mathrm{CF}(x_i)$ of $x_i$ (Line 3).
Here, we consider a protected attribute in a binary setting, grouping it into privileged or unprivileged groups.
Then we audit for fairness violations with respect to the protected attribute, checking whether the outcomes for the corresponding counterfactual pairs change.
When it does, we call these fairness violation instances individual discriminatory instances (IDI). 
There are several methods \cite{angell2018themis, udeshi2018automated, SG-AggarwalESEC/FSE201910.1145/3338906.3338937, fanExpGAICSE22, AFT-zhao-ase2024} to find these discriminatory instances.
For data-driven DPuTs, e.g., DNNs, we also use the \textsc{Themis}~\cite{angell2018themis} implementation by Zhao et al.~\cite{AFT-zhao-ase2024}, \revision{ \toolnameExpGA~\cite{fanExpGAICSE22}, and \toolnameLimi~\cite{LIMI-XiaoISSTA2023-10.1145/3597926.3598099}} to identify and merge these IDIs.
Here, based on the $X_{orig}$, $Y_{orig}$, $\texttt{DPuT}$, and the corresponding IDI search, we can get their corresponding counterfactual input/output $X_{cf}$, $Y_{cf}$ (Line 2-5 in Algorithm \ref{alg:unflocDT}).


\subsection{Curating Relational Dataset}\label{cur-rel-data}
From the $\texttt{DPuT}$ programs (symbolic or black-box), original instances, and counterfactuals (and IDIs), we get two types of pairs in combination with original ($X_{orig}$, $Y_{orig}$) and counterfactual ($X_{cf}$, $Y_{cf}$) data points. 
One set of inputs or data points, even when their protected attributes have been flipped (from privileged to unprivileged or vice versa),  both get the same target outcome from the DPuT. 
We refer to them as fair (or agreement) pairs and assign the relational label $+$ to them.
Another set of inputs or data points, where their protected attributes have been flipped, and where both points do not receive the same target outcome (favorable vs unfavorable) from the predictive model or DPuT.
We refer to them as discriminatory (or disagreeable) pairs, 
and assign the relational label $-$ to them. 
We store these relational labels (fair vs discriminatory) corresponding to each data in $Y_{disc}$ (Line 6-8 in Algorithm \ref{alg:unflocDT}).
And here we create our relational dataset $\mathcal{D}_{\text{R}}$.
$$\mathcal{D}_{\text{R}} = \{ ((x_1, x'_1), L_1), ((x_2, x'_2), L_2), \dots, ((x_n, x'_n), L_n) \}$$
\toolname transforms $\mathcal{D}_{\text{R}}$ into a supervised training dataset $\mathcal{SD}_{\text{m}}$ for the classification of discriminatory pairs from the fair pairs using one of the following curation operator $m \in \{ DCNE, DCVE, DCHE \}$.

$$\mathcal{SD}_{\text{m}} \leftarrow \{(u_i,L_i)\}_{i=1}^n$$ 


\subsubsection{Approach 1: Dataset curation with no extension (DCNE)}
\revision{
For DCNE, only the original instance $x_i$ from each counterfactual pair is retained, 
annotated with the relational label $L_i$ encoding whether the pair $(x_i, x_i')$ 
produced a fair or discriminatory outcome:
$$\mathcal{SD}_{\text{DCNE}} \leftarrow \{(x_i, L_i)\}_{i=1}^n$$
DCNE preserves relational information through labeling rather than explicit pairing: 
$L_i$ encodes the outcome consistency between $x_i$ and its counterfactual $x_i'$, 
so the learned model captures which non-protected feature values trigger sensitivity 
to the protected attribute, without storing counterfactual features directly. The classification label $Y_{orig}$ is dropped.
This avoids introducing potentially unsound feature combinations, which is particularly 
important for black-box DPuTs where counterfactual validity cannot be verified.
}

\begin{algorithm}[t!]
{
    \KwIn{\revision{Program under test }\texttt{DPuT}, \revision{Dataset} $X_{orig}$, \revision{Curation $appr$: `DCNE', `DCVE', or `DCHE'}
    }
    \KwOut{mitigation rules $\mathcal{R}$, \revision{Deferred or fair prediction} $y_{pred}$}
        $Y_{orig}$ $\gets$ getOutcome($X_{orig}$, $\texttt{DPuT}$)
        
        
        \For{$i \leftarrow 1$ \KwTo $|X_{orig}|$}{
          $x'_i \leftarrow \mathrm{CF}(x_i)$\
          {\scriptsize \tcp*{Get corresponding counterfactual input}}
          
          $y'_i \leftarrow f(x'_i)$\
          
          $X_{cf}$.\texttt{add}($x'_i$),\quad $Y_{cf}$.\texttt{add}($y'_i$)\\          
        }

        \ForEach{\((y_i, y'_i) \in (Y_{orig}, Y_{cf})\)}{
            $L_i \leftarrow \mathbf{1}[y_i \neq y'_i]$\
            
            $Y_{disc}$.\texttt{add}($L_i$)
            {\scriptsize \tcp*{Add aggreement/disagreement relational labels}}
        }

          \If{$appr$ == \revision{DCNE}} {
                  $\mathcal{SD}_{\text{\revision{DCNE}}} \leftarrow \{(x_i,L_i)\}_{i=1}^n$
          }
          \ElseIf{$appr$ == \revision{DCVE}}{
            $\mathcal{SD}_{\text{\revision{DCVE}}} \leftarrow \{(x_i,L_i), (x'_i,L_i)\}_{i=1}^n$
          }
          \ElseIf{$appr$ == \revision{DCHE}}{
            $\mathcal{SD}_{\text{\revision{DCHE}}} \leftarrow \{(h(x_i, x'_i),L_i)\}_{i=1}^n$
          }

        $ \mathcal{E},  M  \leftarrow \mathcal{IM}_{train}(S_{appr}, depth_{\max},Leaf_{\max})$\;

        \ForEach{leaf $\lambda$ in $\mathcal{E}$}{
          \If{ isDiscriminatoryLeaf($\lambda$)} {
            $\pi \leftarrow extractRuleForLeaf(\lambda)$
            
            \( \mathcal{R}.\mathtt{add}(\pi) \) \;
          }
        }
        
        debugAndGuardRail($\mathcal{R}$, M)
        {\scriptsize \tcp*{Using rules/explanation to guard rail the model}}
        \ForEach{$x_{pred}$ in $X_{pred}$ }{
            \If{ doesFallUnderDiscriminatoryRule($\mathcal{R}$,$x_{pred}$)  } {
            Deny output
            }
            \Else{
            $y_{pred} \leftarrow DPuT(x_{pred})$
            }
        }
        
    \Return {($\mathcal{R}$, $y_{pred}$)}

\caption{\textsc{\toolname Discrimination identification and verification} 
}
\label{alg:unflocDT}
}
\end{algorithm}


        
        




            

                

                

\subsubsection{Approach 2: Dataset Curation With Vertical Extension (DCVE)}
In this curation technique, both sets of input from each pair $(x_i,  x'_i)$ are used, and the classification label is determined based on their relational label $L_i$ (whether they are from a fair or unfair pair). 
But each element of the pair (original $x_i$ and counterfactual i.e., $x'_i$) is treated as a separate data point, and their corresponding relational labels are $L_i$.
So we have one pair ($(x_i,L_i)$) and another pair ($(x'_i,L_i)$)
In this way our dataset size doubles in terms of rows.
The intuition behind this form of training is that if we add more instances from both fair and unfair pairs, and a better pattern emerges in the classification boundary, it will help the interpretable model distinguish unfair pairs from fair ones.

$$\mathcal{SD}_{\text{DCVE}} \leftarrow \{(x_i,L_i), (x'_i,L_i)\}_{i=1}^n$$ 
where size of the dataset is $N_{DCVE} = 2 \times n$

\subsubsection{Approach 3: Dataset Curation With Horizontal Extension (DCHE)} 
\label{subsec:appr_extend_cols}
For this curation technique, we also take information from both set of input from each pair $(x_i,  x'_i)$, but they are merged into a single row. 
We add columns from both sources ($x_i \in X_{orig}$ and $x'_i \in X_{cf}$) as training features which has different values, dropping one of the columns that has the same values in both sources. 
Let $C = |x_i|$, where $C$ is the number of features we have, for $x_{i}$.  
We can write $x_{i} = (v_{i,1}, ..., v_{i,C})$ and $x'_{i} = (v'_{i,1}, ..., v'_{i,C} )$, where $v_{i,1}$ represent the first feature of the $x_i$ instance. 
And we can have a mask $m_{i,k} = 1[v_{i,k} \neq v'_{i,k}] \in \{0, 1\}$ to drop the features from $x'_{i}$ which are the same as $x'_{i}$.
If we denote the DCHE horizontal feature set as $h(\cdot)$, then
  $$h(x_i, x'_i) = ( v_{i,1}, ..., v_{i,C}, m_{i,1}*v'_{i,1}, ..., m_{i,C} * v'_{i,C} )$$

For $m_{i,k} = 1$, it will add the counterfactual values as they differ, and for $m_{i,k} = 0$, it will drop the identical features.
$$\mathcal{SD}_{\text{DCHE}} \leftarrow \{(h(x_i, x'_i),L_i)\}_{i=1}^n$$ 
And the classification label for each instance is its corresponding relation label $L_i$.
So, we get with relational dataset $\mathcal{SD}_{\text{DCHE}}$.
The intuition behind this form of training is that for each point, we add more information, so that in terms of both original and counterfactual values, the models get more intuition about where the fairness bugs are mostly located.

\subsection{Interpretable Learning and Synthesizing the explanation}
\label{syn-expl}

Given a curated relational dataset $SD_m$ (where $m \in \{DCNE, DCVE, DCHE\}$), we train an interpretable model $\mathcal{M}$ (e.g., a decision tree) to classify fair versus unfair relational pairs (Alg. \ref{alg:unflocDT}, Line 15). Interpretability is essential to capture the feature interactions that localize discriminatory regions within the DPuT. Following \textsc{AFT}~\cite{AFT-zhao-ase2024}, we constrain the model's complexity by limiting the number of leaves. We evaluate $\mathcal{M}$ using standard metrics---Accuracy, Precision, Recall, and ROC-AUC---where True Positives ($TP$) represent correctly identified IDIs, True Negatives $TN$ stands for correctly identified non-IDI, and accuracy measures the fraction of all data points (both fair and unfair) that are correctly classified.
\revision{
Impurity metric depends on the specific tree learner e.g.
ID3 and C4.5 algorithms use cross-entropy, whereas CART decision trees use Gini impurity. 
}

\noindent Once $\mathcal{M}$ achieves sufficient discriminative performance, we extract the path predicates leading to leaves $\lambda$ classified as discriminatory. These paths form a set of explanatory rules $\mathcal{R} = \{\phi_\pi\}$ that localize the specific attribute ranges where discrimination occurs (Alg. \ref{alg:unflocDT}, Lines 16--19). We validate each rule $\phi_\pi$ by subsetting the relational dataset using the rule's conjunction of predicates:
\[
SD_m[\phi_\pi] = \{ (u_i, L_i) \in SD_m \mid \phi_\pi(u_i) = \text{true} \}
\]

For each rule, we compute a few key performance metrics, e.g., impurity, $\text{Confidence}_{leaf}$, and $\text{Coverage}_{leaf}$, to assess the strength of each rule.
$\text{Confidence}_{leaf}$ describes the precision of the corresponding rule, i.e., the fraction of instances captured by the rule $\pi$ that are IDI cases. 
$$
\text{Confidence}_{leaf}(\pi) = \frac{\text{TP}_{\text{leaf}}}{\text{TP}_{\text{leaf}} + \text{FP}_{\text{leaf}}}
$$
Here $\text{TP}_{\text{leaf}}$ stands for the number of actual IDIs that fall under this leaf, and $\text{FP}_{\text{leaf}}$ stands for the number of points that were classified as IDI, but originally was not IDIs.
$$
\text{Coverage}_{leaf} = \frac{\text{TP}_{\text{leaf}}}{ \text{Total \#IDI} }
$$
$\text{Coverage}_{leaf}$ metric finds the fraction of all true discriminatory instances that fall within leaf-rule $\pi$.

\subsection{Mitigation via Guardrail}~\label{mit-gaurd}
\noindent With the extracted rules we have, we add them beside our DPuTs as guardrails to deny output in those stages. When an input is sent to DpuT, instead of sending it directly for prediction, we at first see if the input falls under any of the extracted discriminatory rules ($\mathcal{R}$). If so we refuse to pass it to our DPuT for a prediction and deny output for those datapoints (Alg. \ref{alg:unflocDT}, Lines 20--26). The idea is to abstain from making a decision that could lead to discriminatory behavior.

\section{Experiments}
\label{sec:experiment}
\noindent \textbf{Benchmarks.}\label{benchmark} 
We test our framework \toolname on both symbolic and data-driven programs.


\begin{table*}[!tbh]
\caption{\revision{The dataset and benchmark (symbolic DPuT, left; DNN, right).}}
\centering
\begin{revisionblock}

\begin{minipage}[t]{0.38\textwidth}
\centering
\resizebox{\textwidth}{!}{%
\begin{tabu}{|l|l|l|p{1.1cm}|l|l|}
\hline
DPuT & Dataset & Prot. Feat & \#Inst. & \#Feat. & Source \\\hline
Pg1 & \cite{fairsquare17oopsla} & ethnicity & 30000 & 4 & \cite{fairsquare17oopsla} \\ \hline
Pg2 & \cite{compas-dataset, ZhongNIPS2023GSGAMS-10.5555/3666122.3668598, larson2016we} & sex & 6908 & 4 & \cite{rudin2019stop} \\ \hline
Pg3 & \cite{fairsquare17oopsla} & ethnicity & 30000 & 4 & \cite{Zhang-FairWare22-FairRepair} \\ \hline
Pg4 & \cite{fairsquare17oopsla} & ethnicity & 30000 & 4 & \cite{Zhang-FairWare22-FairRepair} \\ \hline
Pg5  & \cite{compas-dataset, ZhongNIPS2023GSGAMS-10.5555/3666122.3668598, larson2016we} & age & 6908 & 4 & \cite{Rudin2018opt} \\ \hline
Pg6  & \cite{compas-dataset, ZhongNIPS2023GSGAMS-10.5555/3666122.3668598, larson2016we} & age & 6908 & 5 & \cite{Rudin2018opt} \\ \hline
\end{tabu}
}
\end{minipage}%
\hfill
\begin{minipage}[t]{0.60\textwidth}
\centering
\resizebox{\textwidth}{!}{%
\begin{tabu}{|l|l|l|l|l|l|l|l|l|}
\hline
Dataset & Prot. Feat & \#Inst. & \#Feat. & DNN & Source & \texttt{\#Layers} & \texttt{\#Neurons} & Acc (\%) \\\hline
\multirow{12}{*}{\rotatebox[origin=c]{90}{Adult Census}}
 & & \multirow{12}{1.0em}{32,561} & \multirow{12}{1.0em}{13} & AC1  & \cite{kaggle} & 4 & 45  & 85.24 \\ \cline{5-9}
 & & & & AC2  & \cite{kaggle}, \cite{fanExpGAICSE22, udeshi2018automated} & 3 & 121 & 84.70 \\ \cline{5-9}
 & & & & AC3  & \cite{kaggle} & 3 & 71  & 84.52 \\ \cline{5-9}
 & Sex & & & AC4  & \cite{kaggle} & 4 & 221 & 84.86 \\ \cline{5-9}
 & Race & & & AC5  & \cite{kaggle} & 4 & 149 & 85.19 \\ \cline{5-9}
 & Age & & & AC6  & \cite{kaggle} & 4 & 45  & 84.77 \\ \cline{5-9}
 & & & & AC7  & \cite{zhang2020white} & 7 & 145 & 84.85 \\ \cline{5-9}
 & & & & AC8  & \cite{10.1145/3428253, 10.1007/978-3-030-88806-0_15} & 4 & 10  & 82.15 \\ \cline{5-9}
 & & & & AC9  & \cite{10.1145/3428253, 10.1007/978-3-030-88806-0_15} & 6 & 12  & 81.22 \\ \cline{5-9}
 & & & & AC10 & \cite{10.1145/3428253, 10.1007/978-3-030-88806-0_15} & 6 & 20  & 78.56 \\ \cline{5-9}
 & & & & AC11 & \cite{10.1145/3428253, 10.1007/978-3-030-88806-0_15} & 6 & 40  & 79.25 \\ \cline{5-9}
 & & & & AC12 & \cite{10.1145/3428253, 10.1007/978-3-030-88806-0_15} & 11 & 45  & 81.46 \\ \cline{5-9}
\hline
\multirow{8}{*}{\rotatebox[origin=c]{90}{Bank Marketing}}
 & \multirow{8}{4.0em}{Age} & \multirow{8}{4.0em}{45,211} & \multirow{8}{4.0em}{16} & BM1 & \cite{kaggle} & 4 & 97  & 89.20 \\ \cline{5-9}
 & & & & BM2 & \cite{kaggle} & 4 & 65  & 88.76 \\ \cline{5-9}
 & & & & BM3 & \cite{kaggle}, \cite{fanExpGAICSE22, udeshi2018automated} & 3 & 117 & 88.22 \\ \cline{5-9}
 & & & & BM4 & \cite{kaggle} & 5 & 318 & 89.55 \\ \cline{5-9}
 & & & & BM5 & \cite{kaggle} & 4 & 49  & 88.90 \\ \cline{5-9}
 & & & & BM6 & \cite{kaggle} & 4 & 35  & 88.94 \\ \cline{5-9}
 & & & & BM7 & \cite{kaggle} & 4 & 145 & 88.70 \\ \cline{5-9}
 & & & & BM8 & \cite{zhang2020white} & 7 & 141 & 89.20 \\ \cline{5-9}
\hline
\end{tabu}
}
\end{minipage}

\end{revisionblock}
\label{table:prog_benchmark}
\label{table:dataset}
\label{table:ddm-dataset-nn_benchmark}
\end{table*}

\vspace{0.25 em}
\noindent \textit{Symbolic DPuT Benchmarks (PG1–PG6):}
\label{par:bmark-DPuT-desc}
We evaluate our approach on six symbolic DPuTs (PG1-PG6) where the internal decision logic is known, allowing us to validate root-cause explanations against ground truth.
Datasets for PG1, PG3, and PG4 are created following the program listed in Listing~\ref{lst:FairsquarePopulation}, following the work of Albarghouthi et al.~\cite{fairsquare17oopsla}.
\noindent For PG2, PG5, PG6, we use the Compas~\cite{compas-dataset} dataset, we inherit the preprocessed version of the data from \cite{ZhongNIPS2023GSGAMS-10.5555/3666122.3668598, larson2016we}.
\revision{Table~\ref{table:prog_benchmark} (left) shows the details.}

\noindent We evaluate our approach using six symbolic programs (PG1--PG6) representing diverse discriminatory scenarios. PG1, PG3, and PG4 model hiring decisions; PG1~\cite{fairsquare17oopsla} involves indirect discrimination via a proxy (college rank), while PG3 and its supposedly repaired variant PG4~\cite{Zhang-FairWare22-FairRepair} exhibit distinct decision logics. PG2~\cite{rudin2019stop} explicitly incorporates a protected attribute (sex) alongside societal bias proxies like arrest records. Finally, PG5 and PG6~\cite{Rudin2018opt} utilize additive scoring systems where age and prior arrests influence high-risk classification thresholds (see detailed code and logic in ~\cite{Akash-REMI-ISSTA-2026-SupplMat}).

%

\vspace{0.25 em}
\noindent \textit{Neural DPuT Benchmarks.}
To assess the generalization of our approach to black-box models, we employ 20 Deep Neural Networks (DNNs) curated by Biswas et al. \cite{BiswasICSE2023Fairify-10.1109/ICSE48619.2023.00134} from literature~\cite{10.1007/978-3-030-88806-0_15,10.1145/3428253, zhang2020white,udeshi2018automated, fanExpGAICSE22} and Kaggle~\cite{kaggle}. These black-box models encompass various feed-forward architectures with ReLU activations. 
\revision{Table~\ref{table:ddm-dataset-nn_benchmark} (right) shows the characteristics of benchmarks.}
\begin{itemize}[leftmargin=*]
    \item \textit{AC1--AC12:} Trained on the \textit{Adult Census} dataset (32,561 records, 13 attributes) to predict if annual income exceeds \$50,000~\cite{Dua:2019-adult}.
    \item \textit{BM1--BM8:} Trained on the \textit{Bank Marketing} dataset (45,211 entries, 16 features) to predict term deposit subscriptions for a Portuguese bank~\cite{bank_marketing_222}.
\end{itemize}

\vspace{0.25 em}
\noindent \textbf{Technical Details.} 
\toolname was implemented in Python v3.8.20, tensorflow v2.13.0, and scikit-learn v1.3.1
We run all our experiments on an Ubuntu 22.04.5 LTS (jammy) served by an instance type of
c5ad.2xlarge (4 cores, 8 vCPUs, 0 GPUs) in Amazon Web Services (AWS) Elastic Computing (EC2).
We repeat our experiments 10 times and then report the mean and standard deviation.
For RQ5, we run the \toolThemis~\cite{angell2018themis}, \toolnameExpGA~\cite{fanExpGAICSE22}, etc., for 60 minutes to find ID instances in the DNN models.

\vspace{0.25 em}
\noindent \textit{Hyperparameter Selection.}
For decision trees with CART, following AFT~\cite{AFT-zhao-ase2024}, we also set the maximum leaf nodes parameter to 1000 for decision trees. For other models, we follow the same settings.

\noindent \textit{Common Comparison Metrics:}
\label{par:comm-comp-met}
Throughout our experiment section, we use the following metrics. 
The column labeled $Acc$ refers to the accuracy of the model, 
$Prec$ refers to the precision, 
$Rec$ refers to the recall, 
$F1$ refers to the F1 score for classification, which is a harmonic mean of precision and recall, 
$ROC AUC$ is the Area under the curve (AUC) of the receiver-operating characteristic (ROC) score. 
The other relevant metrics will described accordingly in the relevant RQs.

\revision{
\noindent \textbf{Baseline Selection and Scope.}
We select baselines that represent distinct paradigms in fairness testing and mitigation.
\textbf{AFT}~\cite{AFT-zhao-ase2024} is the most directly comparable baseline for the 
\textit{explanation} task, as it also employs a surrogate decision tree to identify 
discriminatory input regions. AFT iteratively generates random inputs within valid bounds, 
trains decision trees on the DPuT's classification labels, and extracts path pairs to guide 
IDI discovery. The key distinction is that AFT trains its trees on the original 
classification task, whereas \toolname trains on relational (fair vs.\ unfair) labels 
derived from counterfactual pairs. We compare both tools on their ability to accurately 
localize discriminatory input regions, using both symbolic and DNN benchmarks.

\noindent \textbf{\toolnameExpGA}~\cite{fanExpGAICSE22} uses explanation-guided genetic 
algorithms to generate IDIs efficiently in a black-box setting. 
\textbf{\toolnameLimi}~\cite{LIMI-XiaoISSTA2023-10.1145/3597926.3598099} generates 
realistic IDIs via latent-space imitation learning.
Both tools focus on point-wise IDI discovery and mitigate unfairness through counterfactual 
data augmentation and retraining --- they do not extract interpretable regional invariants, 
making them unsuitable as explanation baselines.
We therefore compare \toolnameExpGA and \toolnameLimi~against \toolname on the 
\textit{mitigation} task only, using their discovered IDIs as input to \toolname's 
relational pipeline. We exclude \toolThemis, \toolnameExpGA, and \toolnameLimi~from 
symbolic benchmark comparisons, as these tools are designed for black-box models.
}

In this research work, we approach the following research questions:
\begin{enumerate}[start=1,label={\bfseries RQ\arabic*},leftmargin=3em]
\item Which relational dataset curation technique leads to accurate and precise fairness invariants? 
\item Which interpretable algorithms is the best at extracting discriminatory regions? 
\item What are the characteristics of extracted fairness invariants?

%
\item Can we utilize the extracted discriminatory rules as guardrails for mitigating unfairness? 

\item Does \toolname generalize to address individual discrimination in black-box deep neural networks? 

\end{enumerate}


\subsection{ RQ1. Effectiveness of Relational Dataset Curation}

\begin{table*}[!tbh]

\caption{Comparison between different data pre-processing training approaches
}

\centering
\resizebox{0.99\textwidth}{!}{%
\begin{tabu}{|l|l|l|l|l|l|l|l|l|l|l|}
    \hline
        Prog & Appr & Depth & Acc & Prec & Rec & F1 & ROC AUC & $T_{Train}$ & $\#IDI_{Loc}$ \\
        \hline
        
        \multirow{4}{2.5 em}{Pg1 (SMB)} &                 
        most-freq & NA & 0.92 ($\pm$ 0.0) & 0.0 ($\pm$ 0.0) & 0.0 ($\pm$ 0.0) & 0.0 ($\pm$ 0.0) & 0.5 ($\pm$ 0.0) & NA & 0.0 ($\pm$ 0.0) \\
        ~ & AFT~\cite{AFT-zhao-ase2024} & 11.2 ($\pm$ 1.14) & 1.0 ($\pm$ 0.0) & 1.0 ($\pm$ 0.0) & 1.0 ($\pm$ 0.0) & 1.0 ($\pm$ 0.0) & 1.0 ($\pm$ 0.0) & 0.32 ($\pm$ 0.02) & 0.0 ($\pm$ 0.0) \\
        ~ & DCNE & 18.0 ($\pm$ 0.0) & 1.0 ($\pm$ 0.0) & \textbf{0.99} ($\pm$ 0.0) & 0.99 ($\pm$ 0.0) & 0.99 ($\pm$ 0.0) & 1.0 ($\pm$ 0.0) & 0.38 ($\pm$ 0.01) & \textbf{2233.5} ($\pm$ 1.18) \\

        ~ & DCVE & 30.0 ($\pm$ 0.0) & 0.99 ($\pm$ 0.0) & 0.95 ($\pm$ 0.0) & 0.96 ($\pm$ 0.0) & 0.95 ($\pm$ 0.0) & 0.98 ($\pm$ 0.0) & 4.17 ($\pm$ 0.05) & 4329.6 ($\pm$ 2.32) \\         
        
        ~ & DCHE & 17.0 ($\pm$ 0.0) & 1.0 ($\pm$ 0.0) & \textbf{0.99} ($\pm$ 0.0) & 0.99 ($\pm$ 0.0) & 0.99 ($\pm$ 0.0) & 1.0 ($\pm$ 0.0) & 0.37 ($\pm$ 0.01) & \textbf{2231.5} ($\pm$ 1.65) \\         
        \hline
        
        \multirow{3}{3 em}{Pg2 (SMB)} & most-freq & NA & 0.97 ($\pm$ 0.0) & 0.0 ($\pm$ 0.0) & 0.0 ($\pm$ 0.0) & 0.0 ($\pm$ 0.0) & 0.5 ($\pm$ 0.0) & NA & 0.0 ($\pm$ 0.0) \\        
         ~ & AFT~\cite{AFT-zhao-ase2024} & 4.0 ($\pm$ 0.0) & 1.0 ($\pm$ 0.0) & 1.0 ($\pm$ 0.0) & 1.0 ($\pm$ 0.0) & 1.0 ($\pm$ 0.0) & 1.0 ($\pm$ 0.0) & 0.07 ($\pm$ 0.02) & 100.0 ($\pm$ 0.0) \\

        ~ & DCNE &  2.0 ($\pm$ 0.0) & 1.0 ($\pm$ 0.0) & \textbf{1.0} ($\pm$ 0.0) & 1.0 ($\pm$ 0.0) & 1.0 ($\pm$ 0.0) & 1.0 ($\pm$ 0.0) & 0.02 ($\pm$ 0.0) & \textbf{206.0} ($\pm$ 0.0) \\
        ~ & DCVE & 2.0 ($\pm$ 0.0) & 1.0 ($\pm$ 0.0) & \textbf{1.0} ($\pm$ 0.0) & 1.0 ($\pm$ 0.0) & 1.0 ($\pm$ 0.0) & 1.0 ($\pm$ 0.0) & 0.02 ($\pm$ 0.0) & \textbf{412.0} ($\pm$ 0.0) \\
        ~ & DCHE &  2.0 ($\pm$ 0.0) & 1.0 ($\pm$ 0.0) & \textbf{1.0} ($\pm$ 0.0) & 1.0 ($\pm$ 0.0) & 1.0 ($\pm$ 0.0) & 1.0 ($\pm$ 0.0) & 0.02 ($\pm$ 0.0) & \textbf{206.0} ($\pm$ 0.0) \\

        \hline

        \multirow{3}{3 em}{Pg3 (SMB)} & most-freq & NA & 0.97 ($\pm$ 0.0) &0.0 ($\pm$ 0.0) & 0.0 ($\pm$ 0.0) & 0.0 ($\pm$ 0.0) & 0.5 ($\pm$ 0.0) & NA & 0.0 ($\pm$ 0.0) \\

        ~ & AFT~\cite{AFT-zhao-ase2024} & 2.0 ($\pm$ 0.0) & 1.0 ($\pm$ 0.0) & 1.0 ($\pm$ 0.0) & 1.0 ($\pm$ 0.0) & 1.0 ($\pm$ 0.0) & 1.0 ($\pm$ 0.0) & 0.25 ($\pm$ 0.02) & 0.0 ($\pm$ 0.0) \\

        ~ & DCNE & 13.0 ($\pm$ 0.0) & 1.0 ($\pm$ 0.0) & \textbf{1.0} ($\pm$ 0.0) & 1.0 ($\pm$ 0.0) & 1.0 ($\pm$ 0.0) & 1.0 ($\pm$ 0.0) & 0.07 ($\pm$ 0.0) & \textbf{965.5} ($\pm$ 1.58) \\
        
        ~ & DCVE & 24.0 ($\pm$ 0.0) & 1.0 ($\pm$ 0.0) & \textbf{0.97} ($\pm$ 0.0) & 0.97 ($\pm$ 0.0) & 0.97 ($\pm$ 0.0) & 0.98 ($\pm$ 0.0) & 0.94 ($\pm$ 0.02) & 1872.4 ($\pm$ 2.27) \\
        ~ & DCHE & 3.0 ($\pm$ 0.0) & 1.0 ($\pm$ 0.0) & \textbf{1.0} ($\pm$ 0.0) & 1.0 ($\pm$ 0.0) & 1.0 ($\pm$ 0.0) & 1.0 ($\pm$ 0.0) & 0.05 ($\pm$ 0.0) & \textbf{968.0} ($\pm$ 0.0) \\

        \hline
        
        \multirow{3}{3 em}{Pg4 (SMB) } & most-freq & NA & 0.98 ($\pm 0.0$) & 0.0 ($\pm$ 0.0) & 0.0 ($\pm$ 0.0) & 0.0 ($\pm$ 0.0) & 0.5 ($\pm$ 0.0) & NA & 0 ($\pm$ 0.0) \\

        ~ & AFT~\cite{AFT-zhao-ase2024}  & 2.0 ($\pm$ 0.0) & 1.0 ($\pm$ 0.0) & 1.0 ($\pm$ 0.0) & 1.0 ($\pm$ 0.0) & 1.0 ($\pm$ 0.0) & 1.0 ($\pm$ 0.0) & 0.25 ($\pm$ 0.02) & 0.0 ($\pm$ 0.0) \\
        
        ~ & DCNE & 11.0 ($\pm$ 0.0) & 1.0 ($\pm$ 0.0) & \textbf{1.0} ($\pm$ 0.0) & 1.0 ($\pm$ 0.0) & 1.0 ($\pm$ 0.0) & 1.0 ($\pm$ 0.0) & 0.06 ($\pm$ 0.0) & \textbf{580.1} ($\pm$ 1.45) \\
        ~ & DCVE & 21.0 ($\pm$ 0.0) & 1.0 ($\pm$ 0.0) & \textbf{0.97} ($\pm$ 0.0) & 0.98 ($\pm$ 0.0) & 0.98 ($\pm$ 0.0) & 0.99 ($\pm$ 0.0) & 0.5 ($\pm$ 0.01) & 1141.8 ($\pm$ 1.55) \\
        ~ & DCHE & 3.0 ($\pm$ 0.0) & 1.0 ($\pm$ 0.0) & \textbf{1.0} ($\pm$ 0.0) & 1.0 ($\pm$ 0.0) & 1.0 ($\pm$ 0.0) & 1.0 ($\pm$ 0.0) & 0.04 ($\pm$ 0.0) & \textbf{582.0} ($\pm$ 0.0) \\
        \hline

        \multirow{3}{3 em}{Pg5 (Scr)} & most-freq & NA & 0.52 ($\pm$ 0.0) & 0.0 ($\pm$ 0.0) & 0.0 ($\pm$ 0.0) & 0.0 ($\pm$ 0.0) & 0.5 ($\pm$ 0.0) & NA & 0 ($\pm$ 0.0) \\
        ~ & AFT~\cite{AFT-zhao-ase2024}  & 26.3 ($\pm$ 2.0) & 0.75 ($\pm$ 0.01) & \textbf{0.77} ($\pm$ 0.02) & 0.76 ($\pm$ 0.03) & 0.77 ($\pm$ 0.01) & 0.75 ($\pm$ 0.01) & 3.85 ($\pm$ 0.08) & 1178.5 ($\pm$ 152.92) \\
        
        ~ & DCNE & 21.0 ($\pm$ 0.0) & 0.63 ($\pm$ 0.0) & \textbf{0.64} ($\pm$ 0.0) & 0.53 ($\pm$ 0.0) & 0.58 ($\pm$ 0.0) & 0.63 ($\pm$ 0.0) & 2.61 ($\pm$ 0.02) & \textbf{1732.0} ($\pm$ 0.0) \\
        ~ & DCVE & 21.0 ($\pm$ 0.0) & 0.62 ($\pm$ 0.0) & \textbf{0.61} ($\pm$ 0.0) & 0.52 ($\pm$ 0.0) & 0.56 ($\pm$ 0.0) & 0.61 ($\pm$ 0.0) & 3.14 ($\pm$ 0.03) & 3448.2 ($\pm$ 1.14) \\
        ~ & DCHE & 21.0 ($\pm$ 0.0) & 0.63 ($\pm$ 0.0) & \textbf{0.64} ($\pm$ 0.0) & 0.53 ($\pm$ 0.0) & 0.58 ($\pm$ 0.0) & 0.63 ($\pm$ 0.0) & 2.6 ($\pm$ 0.03) & \textbf{1732.0} ($\pm$ 0.0) \\

        \hline

        \multirow{4}{3 em}{Pg6 (Scr)} & most-freq & NA & 0.68 ($\pm$ 0.0) & 0.0 ($\pm$ 0.0) & 0.0 ($\pm$ 0.0) & 0.0 ($\pm$ 0.0) & 0.5 ($\pm$ 0.0) & NA & 0 ($\pm$ 0.0) \\
        ~ & AFT~\cite{AFT-zhao-ase2024} & 3.0 ($\pm$ 0.0) & 1.0 ($\pm$ 0.0) & 1.0 ($\pm$ 0.0) & 1.0 ($\pm$ 0.0) & 1.0 ($\pm$ 0.0) & 1.0 ($\pm$ 0.0) & 0.08 ($\pm$ 0.01) & 100.0 ($\pm$ 0.0) \\
        ~ & DCNE & 6.0 ($\pm$ 0.0) & 1.0 ($\pm$ 0.0) & \textbf{1.0} ($\pm$ 0.0) & 1.0 ($\pm$ 0.0) & 1.0 ($\pm$ 0.0) & 1.0 ($\pm$ 0.0) & 0.03 ($\pm$ 0.0) & \textbf{2216.0} ($\pm$ 0.0) \\
        ~ & DCVE & 5.0 ($\pm$ 0.0) & 1.0 ($\pm$ 0.0) & \textbf{1.0} ($\pm$ 0.0) & 1.0 ($\pm$ 0.0) & 1.0 ($\pm$ 0.0) & 1.0 ($\pm$ 0.0) & 0.03 ($\pm$ 0.0) & \textbf{4432.0} ($\pm$ 0.0) \\
        ~ & DCHE & 6.0 ($\pm$ 0.0) & 1.0 ($\pm$ 0.0) & \textbf{1.0} ($\pm$ 0.0) & 1.0 ($\pm$ 0.0) & 1.0 ($\pm$ 0.0) & 1.0 ($\pm$ 0.0) & 0.03 ($\pm$ 0.0) & \textbf{2216.0} ($\pm$ 0.0) \\

        
        
        
        \hline     
\end{tabu}
}
\label{table:comp_train_appr}
\label{table:rq1_comp_train_appr}
\end{table*}


\noindent \textit{Baseline.}
We consider two approaches:  i) the most frequent labels and ii) AFT~\cite{AFT-zhao-ase2024} as our baseline for localization of discrimination. 
We choose AFT~\cite{AFT-zhao-ase2024} because it is a state-of-the-art individual fairness testing technique that uses a decision tree as a surrogate to infer the individual discriminatory regions. 
Specifically, it uses decision tree path pairs to decide if a subspace is promising for generating individual discriminatory instances. 
The key difference remains that the AFT does not apply a relational input alignment in synthesizing decision trees.
Also, for AFT, instead of multiple iterations of training the decision tree to extract paths for generating tests for individual discrimination (IDs), we stop after one, because our proposed approach trains the interpretable decision tree only once with the corresponding data-curation technique
%

\noindent To determine which relational dataset curation approach is most effective in localizing discrimination, we run experiments with the CART~\cite{Breiman2000ClassificationAR} (decision tree) interpretable model to compare their efficacy. 
The results are summarized in Table \ref{table:comp_train_appr}. 
On the left side of the table, $Prog$ lists relevant DPuTs described in the benchmark.
$T_{Train}$ refers to the time (in seconds) taken to complete the training, 
$\#IDI_{Loc}$ refers to the number of individual discriminatory instances (IDI) localized by the approach. 
Due to the class imbalance, $Prec$,  $Rec$, $F1$, ROC AUC, and $\#IDI_{Loc}$ are terms that are more important to focus on rather than accuracy. 
The most important evaluation metric here is $\#ID_{Loc}$ and $Prec$. 

\noindent From Table \ref{table:comp_train_appr}, we find that our three proposed data-curation techniques localize the largest number of IDIs around 83\% of the time, compared to the considered baselines. 
In more than 80\% of cases, we achieve very good performance metrics for accuracy, precision, recall, F1, and ROC for localizing IDI instances.
Although in one case (Pg5), we can see AFT~\cite{AFT-zhao-ase2024} performs better than the proposed curation techniques. 
AFT identifies more IDIs in PG5, but these AFT-identified IDIs do not necessarily overlap with those considered by the other data curation techniques, due to the nature of the AFT algorithm. 
At each iteration, AFT generates random data within the valid input bounds to find a possible region of interest to generate the IDIs, which is less deterministic and can also generate unrealistic input.
In a few cases, AFT has a better F1 score, but it is worth noting that the decision trees of AFT were trained for an actual classification task of a favorable outcome, whereas our trees were trained for $IDI$ localization, instead of the actual classification tasks.
In most cases, the training approaches with DCNE (no extension) and DCHE (horizontal extension) 
have superior performance in terms of $\#ID$, accuracy, precision, recall, f1-score, all the time.
In terms of the depth of the decision tree, DCHE training needed trees with fewer depths (50\% of cases) to localize the discrimination, whereas the other approaches needed trees with higher depths to achieve similar performance. 
\begin{answerbox}
\textbf{Answer RQ1:} 
Our proposed approaches of creating a relation dataset via pair alignments and training decision tree inference outperform the baselines in 80\% of cases. The horizontal and no-extension alignments are similarly effective at accurately explaining individual discrimination.
\end{answerbox}


\begin{table*}[!tbh]

\caption{Comparison among different interpretable models with training approach with relational dataset.
}
\centering
\resizebox{0.8\textwidth}{!}{%
\begin{tabu}{|l|l|l|l|l|l|l|l|l|l|}
    \hline   
        Prog        & Mod & Acc & Prec & Rec & F1 & ROC AUC & $T_{Tr}$  & $\#ID_{Loc}$ \\
    \hline
\multirow{9}{2.5 em}{Pg1 SMB}  & CART & 1.0 ($\pm$ 0.0) & \textbf{0.99} ($\pm$ 0.0) & 0.99 ($\pm$ 0.0) & 0.99 ($\pm$ 0.0) & 1.0 ($\pm$ 0.0) & 0.38 ($\pm$ 0.01) & \textbf{2233.5} ($\pm$ 1.18) \\

         ~ & GBC & 1.0 ($\pm$ 0.0) & 0.98 ($\pm$ 0.0) & 0.98 ($\pm$ 0.0) & 0.98 ($\pm$ 0.0) & 0.99 ($\pm$ 0.0) & 0.89 ($\pm$ 0.02) & 2196.1 ($\pm$ 9.52) \\
        
        ~ & XGBoost & 1.0 ($\pm$ 0.0) & \textbf{0.98} ($\pm$ 0.0) & 0.99 ($\pm$ 0.0) & 0.98 ($\pm$ 0.0) & 0.99 ($\pm$ 0.0) & 0.15 ($\pm$ 0.19) & \textbf{2233.0} ($\pm$ 0.0) \\
        
        ~ & RuleFit & 0.98 ($\pm$ 0.0) & 0.93 ($\pm$ 0.01) & 0.8 ($\pm$ 0.01) & 0.86 ($\pm$ 0.01) & 0.9 ($\pm$ 0.01) & 64.75 ($\pm$ 1.45) & 1807.7 ($\pm$ 27.51) \\
        
        ~ & FIGS & 0.95 ($\pm$ 0.01) & 0.61 ($\pm$ 0.06) & 0.84 ($\pm$ 0.05) & 0.7 ($\pm$ 0.03) & 0.9 ($\pm$ 0.02) & 0.61 ($\pm$ 0.01) & 1898.8 ($\pm$ 110.81) \\
        
        ~ & C4.5Tree & 1.0 ($\pm$ 0.0) & \textbf{0.99} ($\pm$ 0.0) & 0.99 ($\pm$ 0.0) & 0.99 ($\pm$ 0.0) & 1.0 ($\pm$ 0.0) & 20.36 ($\pm$ 1.33) & \textbf{2232.0} ($\pm$ 4.06) \\
        
        ~ & TaoTree & 0.96 ($\pm$ 0.0) & 0.69 ($\pm$ 0.01) & 0.82 ($\pm$ 0.04) & 0.75 ($\pm$ 0.01) & 0.89 ($\pm$ 0.02) & 2.35 ($\pm$ 0.56) & 1839.4 ($\pm$ 93.1) \\
        
        ~ & OneR & 0.93 ($\pm$ 0.0) & 0.0 ($\pm$ 0.0) & 0.0 ($\pm$ 0.0) & 0.0 ($\pm$ 0.0) & 0.5 ($\pm$ 0.0) & 0.26 ($\pm$ 0.01) & 0.0 ($\pm$ 0.0) \\
        
        ~ & GreedyRuleList & 0.93 ($\pm$ 0.0) & 0.0 ($\pm$ 0.0) & 0.0 ($\pm$ 0.0) & 0.0 ($\pm$ 0.0) & 0.5 ($\pm$ 0.0) & 0.14 ($\pm$ 0.0) & 0.0 ($\pm$ 0.0) \\
         
         \hline
         
\multirow{9}{2.5 em}{Pg2 SMB} & CART & 1.0 ($\pm$ 0.0) & \textbf{1.0} ($\pm$ 0.0) & 1.0 ($\pm$ 0.0) & 1.0 ($\pm$ 0.0) & 1.0 ($\pm$ 0.0) & 0.02 ($\pm$ 0.0) & \textbf{206.0} ($\pm$ 0.0) \\
         ~ & GBC & 1.0 ($\pm$ 0.0) & \textbf{1.0} ($\pm$ 0.0) & 1.0 ($\pm$ 0.0) & 1.0 ($\pm$ 0.0) & 1.0 ($\pm$ 0.0) & 0.05 ($\pm$ 0.0) & \textbf{206.0} ($\pm$ 0.0) \\
        ~ & XGBoost & 1.0 ($\pm$ 0.0) & \textbf{1.0} ($\pm$ 0.0) & 1.0 ($\pm$ 0.0) & 1.0 ($\pm$ 0.0) & 1.0 ($\pm$ 0.0) & 0.03 ($\pm$ 0.0) & \textbf{206.0} ($\pm$ 0.0) \\
        ~ & RuleFit & 1.0 ($\pm$ 0.0) & \textbf{1.0} ($\pm$ 0.0) & 1.0 ($\pm$ 0.0) & 1.0 ($\pm$ 0.0) & 1.0 ($\pm$ 0.0) & 4.97 ($\pm$ 1.38) & \textbf{206.0} ($\pm$ 0.0) \\
        ~ & FIGS & 1.0 ($\pm$ 0.0) & \textbf{1.0} ($\pm$ 0.0) & 1.0 ($\pm$ 0.0) & 1.0 ($\pm$ 0.0) & 1.0 ($\pm$ 0.0) & 0.04 ($\pm$ 0.01) & \textbf{206.0} ($\pm$ 0.0) \\
        ~ & C4.5Tree & 1.0 ($\pm$ 0.0) & \textbf{1.0}($\pm$ 0.0) & 1.0 ($\pm$ 0.0) & 1.0 ($\pm$ 0.0) & 1.0 ($\pm$ 0.0) & 0.21 ($\pm$ 0.02) & \textbf{206.0} ($\pm$ 0.0) \\
        ~ & TaoTree & 1.0 ($\pm$ 0.0) & \textbf{1.0} ($\pm$ 0.0) & 1.0 ($\pm$ 0.0) & 1.0 ($\pm$ 0.0) & 1.0 ($\pm$ 0.0) & 0.07 ($\pm$ 0.0) & \textbf{206.0} ($\pm$ 0.0) \\
        ~ & OneR & 1.0 ($\pm$ 0.0) & 0.94 ($\pm$ 0.0) & 1.0 ($\pm$ 0.0) & 0.97 ($\pm$ 0.0) & 1.0 ($\pm$ 0.0) & 0.05 ($\pm$ 0.0) & \textbf{206.0} ($\pm$ 0.0) \\
        ~ & GreedyRuleList & 1.0 ($\pm$ 0.0) & 0.94 ($\pm$ 0.0) & 1.0 ($\pm$ 0.0) & 0.97 ($\pm$ 0.0) & 1.0 ($\pm$ 0.0) & 0.03 ($\pm$ 0.0) & \textbf{206.0} ($\pm$ 0.0) \\
    \hline
         
\multirow{9}{2.5 em}{Pg3 SMB} & CART & 1.0 ($\pm$ 0.0) & \textbf{1.0} ($\pm$ 0.0) & 1.0 ($\pm$ 0.0) & 1.0 ($\pm$ 0.0) & 1.0 ($\pm$ 0.0) & 0.07 ($\pm$ 0.0) & \textbf{965.5} ($\pm$ 1.58) \\

         ~ & GBC & 1.0 ($\pm$ 0.0) & \textbf{0.99} ($\pm$ 0.0) & 1.0 ($\pm$ 0.0) & 1.0 ($\pm$ 0.0) & 1.0 ($\pm$ 0.0) & 0.29 ($\pm$ 0.0) & \textbf{967.3} ($\pm$ 0.67) \\
        ~ & XGBoost & 1.0 ($\pm$ 0.0) & \textbf{0.98} ($\pm$ 0.0) & 1.0 ($\pm$ 0.0) & 0.99 ($\pm$ 0.0) & 1.0 ($\pm$ 0.0) & 0.09 ($\pm$ 0.03) & \textbf{967.0} ($\pm$ 0.0) \\
        ~ & RuleFit & 1.0 ($\pm$ 0.0) & \textbf{0.99} ($\pm$ 0.0) & 1.0 ($\pm$ 0.0) & 1.0 ($\pm$ 0.0) & 1.0 ($\pm$ 0.0) & 37.19 ($\pm$ 2.15) & \textbf{967.6} ($\pm$ 1.26) \\
        ~ & FIGS & 1.0 ($\pm$ 0.0) & \textbf{1.0} ($\pm$ 0.0) & 1.0 ($\pm$ 0.0) & 1.0 ($\pm$ 0.0) & 1.0 ($\pm$ 0.0) & 0.54 ($\pm$ 0.08) & \textbf{966.3} ($\pm$ 1.25) \\
        ~ & C4.5Tree & 1.0 ($\pm$ 0.0) & \textbf{1.0} ($\pm$ 0.0) & 1.0 ($\pm$ 0.0) & 1.0 ($\pm$ 0.0) & 1.0 ($\pm$ 0.0) & 5.61 ($\pm$ 0.02) & \textbf{966.2} ($\pm$ 1.48) \\
        ~ & TaoTree & 1.0 ($\pm$ 0.0) & \textbf{1.0} ($\pm$ 0.0) & 1.0 ($\pm$ 0.0) & 1.0 ($\pm$ 0.0) & 1.0 ($\pm$ 0.0) & 0.48 ($\pm$ 0.02) & \textbf{966.5} ($\pm$ 1.08) \\
        ~ & OneR & 0.97 ($\pm$ 0.0) & 0.53 ($\pm$ 0.0) & 1.0 ($\pm$ 0.0) & 0.69 ($\pm$ 0.0) & 0.98 ($\pm$ 0.0) & 0.16 ($\pm$ 0.0) & \textbf{968.0} ($\pm$ 0.0) \\
        ~ & GreedyRuleList & 0.97 ($\pm$ 0.0) & 0.53 ($\pm$ 0.0) & 1.0 ($\pm$ 0.0) & 0.69 ($\pm$ 0.0) & 0.98 ($\pm$ 0.0) & 0.08 ($\pm$ 0.0) & \textbf{968.0} ($\pm$ 0.0) \\
         
         \hline
         
\multirow{9}{2.5 em}{Pg4 SMB} & CART & 1.0 ($\pm$ 0.0) & 1.0 ($\pm$ 0.0) & 1.0 ($\pm$ 0.0) & 1.0 ($\pm$ 0.0) & 1.0 ($\pm$ 0.0) & 0.06 ($\pm$ 0.0) & \textbf{580.1} ($\pm$ 1.45) \\

         ~ & GBC & 1.0 ($\pm$ 0.0) & 1.0 ($\pm$ 0.0) & 1.0 ($\pm$ 0.0) & 1.0 ($\pm$ 0.0) & 1.0 ($\pm$ 0.0) & 0.26 ($\pm$ 0.01) & \textbf{581.3} ($\pm$ 0.82) \\
        ~ & XGBoost & 1.0 ($\pm$ 0.0) & 0.97 ($\pm$ 0.0) & 1.0 ($\pm$ 0.0) & 0.98 ($\pm$ 0.0) & 1.0 ($\pm$ 0.0) & 0.08 ($\pm$ 0.01) & \textbf{580.0} ($\pm$ 0.0) \\
        ~ & RuleFit & 1.0 ($\pm$ 0.0) & 0.99 ($\pm$ 0.0) & 1.0 ($\pm$ 0.0) & 1.0 ($\pm$ 0.0) & 1.0 ($\pm$ 0.0) & 42.1 ($\pm$ 3.0) & \textbf{581.8} ($\pm$ 0.42) \\
        ~ & FIGS & 1.0 ($\pm$ 0.0) & 1.0 ($\pm$ 0.0) & 1.0 ($\pm$ 0.0) & 1.0 ($\pm$ 0.0) & 1.0 ($\pm$ 0.0) & 0.44 ($\pm$ 0.05) & \textbf{580.4} ($\pm$ 1.17) \\
        ~ & C4.5Tree & 1.0 ($\pm$ 0.0) & 1.0 ($\pm$ 0.0) & 1.0 ($\pm$ 0.0) & 1.0 ($\pm$ 0.0) & 1.0 ($\pm$ 0.0) & 5.52 ($\pm$ 0.03) & \textbf{580.9} ($\pm$ 1.29) \\
        ~ & TaoTree & 1.0 ($\pm$ 0.0) & 1.0 ($\pm$ 0.0) & 1.0 ($\pm$ 0.0) & 1.0 ($\pm$ 0.0) & 1.0 ($\pm$ 0.0) & 0.4 ($\pm$ 0.02) & \textbf{580.8} ($\pm$ 1.81) \\
        ~ & OneR & 0.98 ($\pm$ 0.0) & 0.0 ($\pm$ 0.0) & 0.0 ($\pm$ 0.0) & 0.0 ($\pm$ 0.0) & 0.5 ($\pm$ 0.0) & 0.16 ($\pm$ 0.0) & 0.0 ($\pm$ 0.0) \\
        ~ & GreedyRuleList & 0.98 ($\pm$ 0.0) & 0.0 ($\pm$ 0.0) & 0.0 ($\pm$ 0.0) & 0.0 ($\pm$ 0.0) & 0.5 ($\pm$ 0.0) & 0.09 ($\pm$ 0.01) & 0.0 ($\pm$ 0.0) \\
         \hline
         
\multirow{9}{2.5 em}{Pg5 Scr}  & CART & 0.63 ($\pm$ 0.0) & \textbf{0.64} ($\pm$ 0.0) & 0.53 ($\pm$ 0.0) & 0.58 ($\pm$ 0.0) & 0.63 ($\pm$ 0.0) & 2.61 ($\pm$ 0.02) & \textbf{1732.0} ($\pm$ 0.0) \\

         ~ & GBC & 0.6 ($\pm$ 0.0) & 0.61 ($\pm$ 0.01) & 0.47 ($\pm$ 0.03) & 0.53 ($\pm$ 0.01) & 0.6 ($\pm$ 0.0) & 0.18 ($\pm$ 0.0) & 1531.6 ($\pm$ 87.62) \\
        ~ & XGBoost & 0.61 ($\pm$ 0.0) & 0.61 ($\pm$ 0.0) & 0.49 ($\pm$ 0.0) & 0.55 ($\pm$ 0.0) & 0.61 ($\pm$ 0.0) & 0.04 ($\pm$ 0.01) & 1624.0 ($\pm$ 0.0) \\
        ~ & RuleFit & 0.53 ($\pm$ 0.0) & 0.58 ($\pm$ 0.21) & 0.02 ($\pm$ 0.01) & 0.03 ($\pm$ 0.02) & 0.5 ($\pm$ 0.0) & 6.36 ($\pm$ 0.64) & 54.8 ($\pm$ 43.07) \\
        ~ & FIGS & 0.63 ($\pm$ 0.0) & \textbf{0.63} ($\pm$ 0.01) & 0.54 ($\pm$ 0.03) & 0.58 ($\pm$ 0.01) & 0.63 ($\pm$ 0.0) & 109.12 ($\pm$ 8.8) & \textbf{1785.7} ($\pm$ 113.57) \\
        ~ & C4.5Tree & 0.63 ($\pm$ 0.0) & 0.65 ($\pm$ 0.01) & 0.5 ($\pm$ 0.02) & 0.57 ($\pm$ 0.01) & 0.63 ($\pm$ 0.0) & 1.75 ($\pm$ 0.06) & 1646.2 ($\pm$ 61.05) \\
        ~ & TaoTree & 0.63 ($\pm$ 0.0) & 0.65 ($\pm$ 0.01) & 0.49 ($\pm$ 0.02) & 0.56 ($\pm$ 0.01) & 0.63 ($\pm$ 0.0) & 4.83 ($\pm$ 0.14) & 1617.5 ($\pm$ 71.01) \\
        ~ & OneR & 0.55 ($\pm$ 0.0) & 0.52 ($\pm$ 0.0) & 0.57 ($\pm$ 0.01) & 0.54 ($\pm$ 0.01) & 0.55 ($\pm$ 0.0) & 0.07 ($\pm$ 0.0) & \textbf{1857.6} ($\pm$ 35.63) \\
        ~ & GreedyRuleList & 0.55 ($\pm$ 0.0) & 0.52 ($\pm$ 0.0) & 0.52 ($\pm$ 0.1) & 0.52 ($\pm$ 0.06) & 0.54 ($\pm$ 0.01) & 0.04 ($\pm$ 0.01) & 1698.3 ($\pm$ 317.8) \\
         \hline
         
\multirow{9}{2.5 em}{Pg6 Scr} & CART & 1.0 ($\pm$ 0.0) & \textbf{1.0} ($\pm$ 0.0) & 1.0 ($\pm$ 0.0) & 1.0 ($\pm$ 0.0) & 1.0 ($\pm$ 0.0) & 0.03 ($\pm$ 0.0) & \textbf{2216.0} ($\pm$ 0.0) \\
    ~ & GBC & 1.0 ($\pm$ 0.0) & \textbf{1.0} ($\pm$ 0.0) & 1.0 ($\pm$ 0.0) & 1.0 ($\pm$ 0.0) & 1.0 ($\pm$ 0.0) & 0.09 ($\pm$ 0.0) & \textbf{2215.6} ($\pm$ 0.52) \\
    ~ & XGBoost & 1.0 ($\pm$ 0.0) & \textbf{1.0} ($\pm$ 0.0) & 1.0 ($\pm$ 0.0) & 1.0 ($\pm$ 0.0) & 1.0 ($\pm$ 0.0) & 0.04 ($\pm$ 0.01) & \textbf{2215.0} ($\pm$ 0.0) \\
    ~ & RuleFit & 1.0 ($\pm$ 0.0) & \textbf{1.0} ($\pm$ 0.0) & 1.0 ($\pm$ 0.0) & 1.0 ($\pm$ 0.0) & 1.0 ($\pm$ 0.0) & 4.46 ($\pm$ 0.38) & \textbf{2215.7} ($\pm$ 0.48) \\
    ~ & FIGS & 1.0 ($\pm$ 0.0) & \textbf{1.0} ($\pm$ 0.0) & 1.0 ($\pm$ 0.0) & 1.0 ($\pm$ 0.0) & 1.0 ($\pm$ 0.0) & 0.14 ($\pm$ 0.02) & \textbf{2215.9} ($\pm$ 0.32) \\
    ~ & C4.5Tree & 1.0 ($\pm$ 0.0) &\textbf{1.0} ($\pm$ 0.0) & 1.0 ($\pm$ 0.0) & 1.0 ($\pm$ 0.0) & 1.0 ($\pm$ 0.0) & 0.53 ($\pm$ 0.03) & \textbf{2215.8} ($\pm$ 0.42) \\
    ~ & TaoTree & 1.0 ($\pm$ 0.0) & \textbf{1.0} ($\pm$ 0.0) & 1.0 ($\pm$ 0.0) & 1.0 ($\pm$ 0.0) & 1.0 ($\pm$ 0.0) & 0.22 ($\pm$ 0.01) & \textbf{2215.7} ($\pm$ 0.48) \\
    ~ & OneR & 0.81 ($\pm$ 0.0) & 0.63 ($\pm$ 0.0) & 1.0 ($\pm$ 0.0) & 0.78 ($\pm$ 0.0) & 0.86 ($\pm$ 0.0) & 0.06 ($\pm$ 0.0) & \textbf{2216.0} ($\pm$ 0.0) \\
    ~ & GreedyRuleList & 0.81 ($\pm$ 0.0) & 0.63 ($\pm$ 0.0) & 1.0 ($\pm$ 0.0) & 0.78 ($\pm$ 0.0) & 0.86 ($\pm$ 0.0) & 0.04 ($\pm$ 0.0) & \textbf{2216.0} ($\pm$ 0.0) \\
                  
    \hline
\end{tabu}
}
\label{table:RQ2_comp_ip_models_orig_flip}
\end{table*}


\subsection{RQ2. Performance of Interpretable Algorithms in Inferring Fairness Invariants}
We evaluate nine tree-based algorithms to identify the optimal learner for fairness invariant synthesis: CART \cite{Breiman2000ClassificationAR}, GBC \cite{friedman2001greedyGBM}, XGB \cite{Chen:2016:XST:2939672.2939785}, C4.5 \cite{C4.5-tree-Quinlan}, TaoTree \cite{PerpinanTaoTree-NEURIPS2018_185c29dc}, RuleFit \cite{rulefit-friedman-2008}, FIGS \cite{FIGS-Tan-2025}, OneR \cite{Holte1993VerySC}, and GreedyRuleList \cite{imodels/rule_list/greedy_rule_list}. Table~\ref{table:RQ2_comp_ip_models_orig_flip} summarizes the training time ($T_{Tr}$) and the number of localized IDIs ($\#IDI_{Loc}$) across symbolic DPuTs.

\noindent For \texttt{PG1}, C4.5, CART, and XGB (using DCNE) localized over 2,230 IDIs. While XGB was the fastest, its ensemble nature limits direct interpretability. In \texttt{PG3} and \texttt{PG4}, all major learners showed comparable performance; however, FIGS and TaoTree provided the shortest training times. RuleFit achieved slightly higher localization in PG3, but at the cost of training latency and precision. In \texttt{PG5}, which exhibits low separability due to its probabilistic Bernoulli logic, OneR localized the most IDIs (1,857) but with the lowest precision. FIGS and CART provided a superior balance of precision and localization. For \texttt{PG2} and \texttt{PG6}, simpler decision logic and smaller datasets led to high performance across all algorithms.


\noindent Overall, \textit{C4.5}, \textit{FIGS}, and \textit{CART} emerged as the top performers, outperforming other methods in 83\% of cases. While XGB and GBC are computationally efficient, they lack the inherent interpretability required for precise bug localization. Conversely, OneR and GreedyRuleList frequently underperformed, resulting in low precision and F1 scores.

\begin{answerbox}
\textbf{Answer RQ2:} C4.5, FIGS, and CART are the most effective for precisely inferring fairness invariants, achieving top-tier performance in 83\% of benchmarks.
\end{answerbox}


\begin{table*}[!tbh]

\caption{Comparison between the strength of the extracted rules of using proposed relational dataset (RQ3) 
}

\centering
\resizebox{0.99\textwidth}{!}{%
\begin{tabu}{|l|l|l|l|l|l|l|l|l|l|l|l|}
    \hline
    
        Prog & Appr & Depth & $Rule.Len$ & Imp & Conf & Tot.Disc.Cov & Tot.$\#IDI_{Loc}$ & Bst.Lf.Rule.Len & Bst.Lf.Disc.Cov & Bst.Lf.Conf &  Bst.Lf.$\#IDI_{Loc}$ 
        \\ \hline
        
        \multirow{3}{3 em}{Pg1 SMB} & DCNE & 18.0 ($\pm$ 0.0) & 11.56 ($\pm$ 2.77) & 0.0 ($\pm$ 0.0) & 0.97 ($\pm$ 0.09) & \textbf{0.98} ($\pm$ 0.0) &\textbf{2233.5} ($\pm$ 1.18) & 16.0 ($\pm$ 0.0) & 0.13 ($\pm$ 0.0) & 1.0 ($\pm$ 0.0) & \textbf{301.0} ($\pm$ 0.0)   \\
        
        ~ & DCHE & 17.0 ($\pm$ 0.0) & 10.67 ($\pm$ 2.69) & 0.0 ($\pm$ 0.0) & 0.98 ($\pm$ 0.09) & \textbf{0.98} ($\pm$ 0.0) & \textbf{2231.5} ($\pm$ 1.65) & 15.0 ($\pm$ 0.0) & 0.13 ($\pm$ 0.0) & 1.0 ($\pm$ 0.0) & \textbf{301.0} ($\pm$ 0.0) \\
        
        ~ & \toolAFTWCite & 11.0 ($\pm$ 0.0) & 7.23 ($\pm$ 1.79) & 0.0 ($\pm$ 0.0) & 1.0 ($\pm$ 0.0) & 0.0 ($\pm$ 0.0) & 0.0 ($\pm$ 0.0) &  3.4 ($\pm$ 0.52) & 0 ($\pm$ 0.0) & 1.0 ($\pm$ 0.0) & 0 ($\pm$ 0.0)\\
        
        \hline
        \multirow{3}{3 em}{Pg2 SMB } &DCNE & 2.0 ($\pm$ 0.0) & 2.0 ($\pm$ 0.0) & 0.0 ($\pm$ 0.0) & 1.0 ($\pm$ 0.0) & \textbf{1.0} ($\pm$ 0.0) & \textbf{206.0} ($\pm$ 0.0) & 2.0 ($\pm$ 0.0) & 1.0 ($\pm$ 0.0) & 1.0 ($\pm$ 0.0) & \textbf{206.0} ($\pm$ 0.0) \\
        
        ~ & DCHE & 2.0 ($\pm$ 0.0) & 2.0 ($\pm$ 0.0) & 0.0 ($\pm$ 0.0) & 1.0 ($\pm$ 0.0) & \textbf{1.0} ($\pm$ 0.0) & \textbf{206.0} ($\pm$ 0.0) & 2.0 ($\pm$ 0.0) & 1.0 ($\pm$ 0.0) & 1.0 ($\pm$ 0.0) & \textbf{206.0} ($\pm$ 0.0) \\
        
        ~ & \toolAFTWCite & 4.0 ($\pm$ 0.0) & 2.67 ($\pm$ 1.53) & 0.0 ($\pm$ 0.0) & 1.0 ($\pm$ 0.0) & 1.0 ($\pm$ 0.0) & 100.0 ($\pm$ 0.0) & 4 ($\pm$ 0.0) & 1.0 ($\pm$ 0.0) & 1.0 ($\pm$ 0.0) & 100.0 ($\pm$ 0.0) \\
        \hline
        \multirow{3}{3 em}{Pg3 SMB} & DCNE & 13.0 ($\pm$ 0.0) & 9.4 ($\pm$ 2.5) & 0.0 ($\pm$ 0.0) & 0.99 ($\pm$ 0.01) & \textbf{0.99} ($\pm$ 0.1) & \textbf{965.5} ($\pm$ 1.58) & 5.0 ($\pm$ 0.0) & 0.89 ($\pm$ 0.0) & 1.0 ($\pm$ 0.0) & 863.0 ($\pm$ 0.0) \\
        
        ~ & DCHE & \textbf{3.0} ($\pm$ 0.0) & 3.0 ($\pm$ 0.0) & 0.0 ($\pm$ 0.0) & 1.0 ($\pm$ 0.0) & \textbf{1.0} ($\pm$ 0.0) & \textbf{968.0} ($\pm$ 0.0) & 3.0 ($\pm$ 0.0) & 1.0 ($\pm$ 0.0) & 1.0 ($\pm$ 0.0) & \textbf{968.0} ($\pm$ 0.0) \\        
        ~ & \toolAFTWCite & 2.0 ($\pm$ 0.0) & 1.5 ($\pm$ 0.71) & 0.0 ($\pm$ 0.0) & 1.0 ($\pm$ 0.0) &  0.0 ($\pm$ 0.0) & 0 ($\pm$ 0.0) & 1 ($\pm$ 0.0) & 0 ($\pm$ 0.0) & 1.0  ($\pm$ 0.0) & 0 ($\pm$ 0.0) \\
        \hline
        
        \multirow{3}{3 em}{Pg4 (SMB) } & DCNE & 11.0 ($\pm$ 0.0) & 8.43 ($\pm$ 2.07) & 0.0 ($\pm$ 0.0) & 1.0 ($\pm$ 0.01) & \textbf{0.99} ($\pm$ 0.0) & \textbf{580.1} ($\pm$ 1.45) & 5.0 ($\pm$ 0.0) & 0.89 ($\pm$ 0.0) & 1.0 ($\pm$ 0.0) & 516.0 ($\pm$ 0.0) \\
        
        ~ & DCHE & \textbf{3.0} ($\pm$ 0.0) & 3.0 ($\pm$ 0.0) & 0.0 ($\pm$ 0.0) & 1.0 ($\pm$ 0.0) & \textbf{1.0} ($\pm$ 0.0) & \textbf{582.0} ($\pm$ 0.0) & 3.0 ($\pm$ 0.0) & 1.0 ($\pm$ 0.0) & 1.0 ($\pm$ 0.0) & \textbf{582.0} ($\pm$ 0.0) \\
        
        ~ & \toolAFTWCite & 2.0 ($\pm$ 0.0) & 1.5 ($\pm$ 0.71) & 0.0 ($\pm$ 0.0) & 1.0 ($\pm$ 0.0) &  0 ($\pm$ 0.0)& 0 ($\pm$ 0.0)& 1 ($\pm$ 0.0) & 0 ($\pm$ 0.0) & 1.0 ($\pm$ 0.0) & 0.0 ($\pm$ 0.0) \\
        \hline
        \multirow{3}{3 em}{Pg5 (Scr)} & DCNE & 21.0 ($\pm$ 0.0) & 12.75 ($\pm$ 3.47) & 0.2 ($\pm$ 0.23) & 0.77 ($\pm$ 0.2) & 0.54 ($\pm$ 0.0) & 1732.0 ($\pm$ 0.0) & 9.0 ($\pm$ 0.0) & 0.02 ($\pm$ 0.0) & 0.5 ($\pm$ 0.0) & 63.0 ($\pm$ 0.0) \\
        ~ & DCHE & 21.0 ($\pm$ 0.0) & 12.79 ($\pm$ 3.49) & 0.2 ($\pm$ 0.23) & 0.77 ($\pm$ 0.2) & 0.54 ($\pm$ 0.0) & 1732.0 ($\pm$ 0.0) & 9.0 ($\pm$ 0.0) & 0.02 ($\pm$ 0.0) & 0.5 ($\pm$ 0.0) & 63.0 ($\pm$ 0.0) \\
        
        ~ & \toolAFTWCite & 25.0 ($\pm$ 0.0) & 13.61 ($\pm$ 3.7) & 0.27 ($\pm$ 0.41) & 0.9 ($\pm$ 0.15) & \textbf{0.9} ($\pm$ 0.0) & \textbf{1843} ($\pm$ 0.0) & 4 ($\pm$ 0.0) & 0.49 ($\pm$ 0.0) & 1.0 ($\pm$ 0.0) & \textbf{999} ($\pm$ 0.0) \\
        \hline
        
        \multirow{2}{3 em}{Pg6 (Scr)} & DCNE & 6.0 ($\pm$ 0.0) & 4.6 ($\pm$ 1.14) & 0.0 ($\pm$ 0.0) & 1.0 ($\pm$ 0.0) & \textbf{1.0} ($\pm$ 0.0) & \textbf{2216.0} ($\pm$ 0.0) & 3.0 ($\pm$ 0.0) & 0.62 ($\pm$ 0.0) & 1.0 ($\pm$ 0.0) & \textbf{1378.0} ($\pm$ 0.0) \\        
        
        ~ & DCHE & 6.0 ($\pm$ 0.0) & 4.6 ($\pm$ 1.14) & 0.0 ($\pm$ 0.0) & 1.0 ($\pm$ 0.0) & \textbf{1.0} ($\pm$ 0.0) & \textbf{2216.0} ($\pm$ 0.0) & 3.0 ($\pm$ 0.0) & 0.62 ($\pm$ 0.0) & \textbf{1.0} ($\pm$ 0.0) & \textbf{1378.0} ($\pm$ 0.0) \\
        ~ & \toolAFTWCite & 3.0 ($\pm$ 0.0) & 2.0 ($\pm$ 0.0) & 0.0 ($\pm$ 0.0) & 1.0 ($\pm$ 0.0) & 1.0 ($\pm$ 0.0) & 100.0 ($\pm$ 0.0) & 3 ($\pm$ 0.0) & 1.0 ($\pm$ 0.0) & 1.0 ($\pm$ 0.0) & 100.0 ($\pm$ 0.0) \\
        \hline

\end{tabu}
    }
\label{table:comp-rules-gt-root-cause-trn-app-p2}
\end{table*}

%


\subsection{ RQ3. Various Characteristics of the Relational Explanation Models.}~\label{subsection:remi-rules-rq}
For symbolic programs where the root cause of discrimination is verifiable, we measure how closely \toolname's extracted predicates align with the actual discriminatory conditions.
%
%
Table~\ref{table:comp-rules-gt-root-cause-trn-app-p2} summarizes these findings using the following metrics:
\begin{itemize}[leftmargin=*]
    \item \textbf{Rule Length ($Rul.Len$):} Average number of predicates in rules reaching a discriminatory leaf.
    \item \textbf{Impurity ($Imp$):} Average node impurity across all discriminatory leaves.
    \item \textbf{Confidence ($Conf$):} Average precision of rules in identifying IDIs.
    \item \textbf{Coverage ($Tot.Disc.Cov$ / $Bst.Lf.Disc.Cov$):} Total IDIs captured by all rules versus the single most representative leaf rule.
\end{itemize}

\noindent In PG1 (Hiring), both DCNE and DCHE trees hit zero impurity and high confidence around 0.97, with $Tot.Disc.Cov$ around 0.98 and locates around 2233 IDIs.
The single best leaf rule (out of a total 95 rules) covers around 13\% of the discriminatory instances considered (301 out of 2235).
Average rule lengths are longer, ranging from 10 to 12 literals, indicating that the unsafe region is a narrow area under several tight predicate conditions.
AFT~\cite{AFT-zhao-ase2024} does not find any IDIs for PG1 in the current settings, so we do not have any effective rule to compare with.
For PG1, our framework discovers a compact discriminatory rule that matches the discrimination encoded in the DPuT, where the outcomes are based on ethnicity. 
After simplification, the best leaf rule becomes
\texttt{\{ 14.6 < col-rank $\le$ 17.25 $\land$ 11.99 < y-exp $\le$ 14.84 $\land$ cf-ethnicity > 10.5 $\land$ 17.85< cf-col-rank $\le$ 22.95 \}}
which isolates a narrow region of (col-rank, y-exp) space when the model becomes significantly discriminatory with respect to the protected attribute (ethnicity).
The rule recovers the root cause of the discrimination lies in a narrow college rank, mid-experience band 
(14.6< col-rank $\le$ 17.25, 11.99 < y-exp $\le$ 14.84) for a favorable ethnicity.
Now, flipping the ethnicity to an unfavorable group increases the counterfactual college rank 
(17.85 < cf-col-rank $\le$ 22.95) and flips the hiring decision. 
The rule explicitly captures the ethnicity to college rank to hire pathway, demonstrating the match with the ground truth level root cause of the discrimination of the DPuT PG1.

\noindent For PG2 (Recidivism), \toolname achieves 1.0 confidence and 100\% coverage (206 IDIs) with a simple two-literal rule: \texttt{{age $\le$ 20.5 $\land$ priors-count $\le$ 3.5}}. This matches the ground truth logic regarding gender-based discrimination in young defendants. AFT~\cite{AFT-zhao-ase2024} localizes only 100 IDIs using a similar predicate structure.


\noindent For PG3, \toolname achieves 0.99 confidence and 100\% coverage (968 IDIs), identifying unfair behavior when \texttt{y-exp < 15} and \texttt{col-rank} differ between groups. For PG4 (the repaired variant of the hiring algorithm), \toolname maintains 0.99 confidence and 100\% coverage (582 IDIs), revealing that discriminatory behavior persists within specific \texttt{col-rank} and \texttt{y-exp} ranges. AFT~\cite{AFT-zhao-ase2024} failed to identify IDIs for either program.


\noindent In PG5, \toolname achieves 0.77 confidence and 53\% total coverage (1,732 instances). The extracted rules correctly identify the age bracket (\texttt{20.5 < age $\le$ 21.5}) as the primary discriminatory driver. In PG 6, \toolname achieves 100\% confidence and 100\% coverage. A single extracted rule---incorporating \texttt{age}, \texttt{priors-count}, and \texttt{juv-misd-count}---localizes 62\% (1,378) of total IDIs. \toolname significantly outperforms AFT, which localizes only 100 instances.


\noindent Overall, \toolname precisely characterizes the root cause of discrimination in 83\% of cases.
In all cases, except PG5, \toolname significantly outperforms the state-of-the-art baseline, AFT~\cite{AFT-zhao-ase2024}. Specifically, in 50\% of cases (PG1, PG2, PG4), AFT fails to find any ID instances, rendering its rules inapplicable. This underscores the necessity of relational explanation, as our approach provides greater applicability where standard IDI search methods fail. 

\begin{answerbox}
\textbf{Answer RQ3:} \toolname outperformed the state-of-the-art technique AFT~\cite{AFT-zhao-ase2024} in 83\% of cases. In more than 66\% of cases, \toolname identified all the discrimination regions from the symbolic programs. 
%
\end{answerbox}

\subsection{RQ4. Performance of Bias Mitigation via Fairness Invariant Guardrails}

A common mitigation approach involves retraining the model with individual discrimination instances (IDIs) so it learns to avoid discrimination in local regions~\cite{AFT-zhao-ase2024,fanExpGAICSE22,zhang2020white,udeshi2018automated}. However, our framework provides rules that define the discriminatory region and its root causes.
These rules act as guardrails: when an input falls within a localized discriminatory region, the system can delay inference for human review. This ensures the model maintains relational consistency and reduces errors in real-time deployment.

%


\begin{wraptable}{r}{8cm}
\caption{Comparing performance during Mitigation}
\centering
\resizebox{0.5\textwidth}{!}{%
\begin{tabu}{|l|l|l|l|l|l|l|l|l|}
    \hline
        Prog & Appr & Acc & Prec & Rec & F1 & ROC AUC & $\#IDI$ & $\#FP_{IDI}$  \\
        \hline
        \multirow{4}{2 em}{Pg1 SMB} 
        & Ini. Pg & 1.00 & 1.00 & 1.00 & 1.00 & 1.00 & 2250 & -\\
        ~ & DT w IDI & 0.99 & 0.99 & 0.99 & 0.99 & 0.99 & 2430 & - \\
        ~ & Ini.Pg + \toolname & 1.00 & 1.00 & 1.00 & 1.00 & 1.00  & \textbf{2} & 2670\\
        \hline

        \multirow{4}{2 em}{Pg2 SMB} & Init. Pg & 1.00 & 1.00 & 1.00 & 1.00 & 1.00 &  206 & - \\
        ~ & DT w IDI & 1.00 & 1.00 & 1.00 & 1.00 & 1.00 & 206 & - \\     
        ~ & Ini.Pg + \toolname & 1.00 & 1.00 & 1.00 & 1.00 & 1.00 & \textbf{0} & \textbf{0}\\
        \hline
        \multirow{4}{2 em}{Pg3 SMB} & Init. Pg &1.00 & 1.00 & 1.00 & 1.00 & 1.00 &  968 & - \\
        ~ & DT w IDI & 1.00 & 1.00 & 1.00 & 1.00 & 1.00 & 968 & - \\    
        ~ & Ini.Pg + \toolname & 1.00 & 1.00 & 1.00 & 1.00 & 1.00 & \textbf{0} & 8\\
                   
        \hline
        \multirow{4}{2 em}{Pg4 SMB} & Init. Pg & 1.00 & 1.00 & 1.00 & 1.00 & 1.00 &  582 & - \\
        ~ & DT w IDI & 1.00 & 1.00 & 1.00 & 1.00 & 1.00 & 582 & -\\   
        ~ & Ini.Pg + \toolname & 1.00 & 1.00 & 1.00 & 1.00 & 1.00 & \textbf{0} & 5\\
        \hline

        \multirow{4}{2 em}{Pg5 SC} & Init. Pg & 1.00 & 1.00 & 1.00 & 1.00 & 1.00 &  3286 & - \\
        ~ & DT w IDI & 0.68 & 0.61 & 0.79 & 0.69 & 0.69  & 2895 & -\\      
        ~ & Ini.Pg + \toolname & 1.00 & 1.00 & 1.00 & 1.00 & 1.00 & \textbf{1338} & 1567\\ 
        \hline
        
        \multirow{4}{2 em}{Pg6 SC} & Init. Pg & 1.00 & 1.00 & 1.00 & 1.00 & 1.00 & 2216 & - \\
        ~ & DT w IDI & 1.00 & 1.00 & 1.00 & 1.00 & 1.00  & 2216 & -\\   
        ~ & Ini.Pg + \toolname & 1.00 & 1.00 & 1.00 & 1.00 & 1.00 & \textbf{0} & 32\\
        \hline
        
\end{tabu}
}
\label{table:comp_mitigation}
\end{wraptable}



\noindent Table \ref{table:comp_mitigation} reports the mitigation results across three program versions, focusing on the number of remaining IDIs ($\#IDI$) and the number of non-discriminatory points incorrectly flagged as IDIs ($FP_{IDI}$).
Applying our rules as guardrails reduced $\#IDI$ in 100\% of cases, demonstrating the strong efficacy of the rule-based approach. Performance metrics remained consistently close to those of the original DPuTs. \toolname identifies non-discriminatory points as IDIs in a few instances ($FP_{IDI} > 0$). This occurs when discriminatory and non-discriminatory points share similar attributes or reside across a narrow boundary that the decision tree's hyper-rectangular separation struggles to classify.

\begin{answerbox}
\textbf{Answer RQ4:} The rule-based guardrails reduced the number of individual discriminatory instances from 2,250 to 2 in one case and from hundreds/thousands of instances to 0 in four cases (out of 6). 
\end{answerbox}

\subsection{RQ5. Generalization of \toolname to Black-Box Neural Networks (DNNs).}

Once we establish the precision and usefulness of \toolname relative to verifiable symbolic programs, we apply it to black-box deep neural networks to study the generalization. 
%
\revision{Specifically, we evaluate (a) whether \toolname can perform rule inference, and (b) whether the 
extracted rules reduce discriminatory decisions while preserving the model's original 
classification utility.}

%

\begin{table*}[!tbh]

\caption{\revision{DNN rule extraction results using ExpGA~\cite{fanExpGAICSE22} as the IDI source under DCNE curation strategy (RQ5).}}
\centering
\begin{revisionblock}
\resizebox{0.9\textwidth}{!}{%
\begin{tabu}{|c|c|c|c|c|c|c|c|c|c|c|c|}
    \hline    
    Model & Depth & $Rule.Len$ & $Tot.Rules$ & Imp & Conf & Tot.Disc.Cov & Tot.$\#IDI_{Loc}$ & Bst.Lf.Rule.Len & Bst.Lf.Disc.Cov & Bst.Lf.Conf & Bst.Lf.$\#IDI_{Loc}$ \\
    \hline

    AC1 & 35.0 ($\pm$ 0.0) & 18.77 ($\pm$ 7.08) & 449 & 0.1 ($\pm$ 0.17) & 0.85 ($\pm$ 0.18) & 0.69 & 11496 & 4 & 0.5 & 1.0 & 7793 \\
    \hline
    AC2 & 24.0 ($\pm$ 0.0) & 14.17 ($\pm$ 3.98) & 467 & 0.12 ($\pm$ 0.18) & 0.85 ($\pm$ 0.18) & 0.77 & 15583 & 7 & 0.43 & 1.0 & 8287 \\
    \hline
    AC3 & 29.0 ($\pm$ 0.0) & 15.54 ($\pm$ 4.73) & 439 & 0.11 ($\pm$ 0.18) & 0.83 ($\pm$ 0.19) & 0.78 & 20531 & 5 & 0.66 & 1.0 & 16291 \\
    \hline
    AC4 & 24.0 ($\pm$ 0.0) & 14.45 ($\pm$ 3.91) & 436 & 0.09 ($\pm$ 0.16) & 0.87 ($\pm$ 0.17) & 0.7 & 13864 & 2 & 0.58 & 1.0 & 10738 \\
    \hline
    AC5 & 30.0 ($\pm$ 0.0) & 15.47 ($\pm$ 4.51) & 464 & 0.11 ($\pm$ 0.18) & 0.85 ($\pm$ 0.19) & 0.76 & 18223 & 5 & 0.64 & 1.0 & 14384 \\
    \hline
    AC6 & 30.0 ($\pm$ 0.0) & 16.35 ($\pm$ 4.83) & 455 & 0.1 ($\pm$ 0.17) & 0.85 ($\pm$ 0.19) & 0.53 & 5333 & 2 & 0.23 & 1.0 & 2215 \\
    \hline
    AC7 & 25.0 ($\pm$ 0.0) & 14.65 ($\pm$ 3.98) & 469 & 0.11 ($\pm$ 0.17) & 0.85 ($\pm$ 0.18) & 0.68 & 8942 & 2 & 0.17 & 0.99 & 2095 \\
    \hline
    AC8 & 30.0 ($\pm$ 0.0) & 14.98 ($\pm$ 5.05) & 459 & 0.1 ($\pm$ 0.18) & 0.84 ($\pm$ 0.19) & 0.6 & 7231 & 9 & 0.34 & 1.0 & 3845 \\
    \hline
    AC9 & 30.0 ($\pm$ 0.0) & 16.96 ($\pm$ 4.81) & 449 & 0.11 ($\pm$ 0.18) & 0.83 ($\pm$ 0.19) & 0.41 & 3729 & 2 & 0.1 & 1.0 & 851 \\
    \hline
    AC10 & 24.0 ($\pm$ 0.0) & 14.39 ($\pm$ 3.68) & 441 & 0.11 ($\pm$ 0.18) & 0.84 ($\pm$ 0.19) & 0.84 & 23704 & 5 & 0.7 & 1.0 & 18997 \\
    \hline
    AC11 & 29.0 ($\pm$ 0.0) & 16.17 ($\pm$ 4.68) & 457 & 0.11 ($\pm$ 0.17) & 0.85 ($\pm$ 0.17) & 0.8 & 20979 & 3 & 0.66 & 1.0 & 16368 \\
    \hline
    AC12 & 33.0 ($\pm$ 0.0) & 17.37 ($\pm$ 5.64) & 443 & 0.11 ($\pm$ 0.18) & 0.84 ($\pm$ 0.18) & 0.75 & 14825 & 2 & 0.57 & 1.0 & 10649 \\
    \hline

\end{tabu}
}
\end{revisionblock}
\label{tab:rule-extraction-expga-main-paper}
\end{table*}

\revision{
\noindent\textbf{Localization via Rule Extraction for DNNs.}
Table~\ref{tab:rule-extraction-expga-main-paper} reports rule extraction results for the 12 Adult Census DNN benchmarks using \toolnameExpGA-generated IDIs~\cite{fanExpGAICSE22}.
\toolname produces interpretable trees of moderate depth (24--35), 
yielding 430--500 rules with average lengths of 14--18 conditions.
Extracted rules achieve high average confidence (0.83--0.88), indicating reliable 
mapping between input features and relational fairness labels.
Discrimination coverage ranges from 41--84\%, confirming that the rules capture a 
substantial majority of the discovered discriminatory input space.
}

%

\begin{table*}[t]
\centering
\scriptsize
\caption{\revision{Comparison of mitigation techniques using IDIs generated by black-box DNN fairness testing tools. 
}}
\label{tab:rq5_dnn_compact}
\begin{revisionblock}
{\renewcommand{\arraystretch}{0.86}
\resizebox{0.98\textwidth}{!}{%
\begin{tabular}{@{}c@{\hspace{0.8em}}c@{}}
\begin{tabular}[t]{l l | r r | r r r r}
\toprule
Model & Mitigation & Acc & F1 & $\#\text{IDI}_\text{bef}$ & $\#\text{IDI}_\text{aft}$ & Red.\ (\%) & $\#FP_{IDI}$ \\
\midrule
  \multirow{9}{*}{AC1} & \toolThemisWCite & 0.85 & 0.66 & 7{,}173 & 7{,}571 & -5.55 & -- \\
   & \toolname & 0.85 & 0.64 & 7{,}173 & \textbf{4{,}111} & 42.69 & 1{,}323 \\
   & \toolThemisWCite + \toolname & 0.85 & 0.66 & 7{,}571 & \textbf{4{,}370} & 42.28 & 1{,}208 \\
\cmidrule(lr){2-8}
   &  \toolExpGAWCite & 0.84 & 0.61 & 26{,}647 & 15{,}687 & 41.13 & -- \\
   & \toolname & 0.85 & 0.64 & 26{,}647 & \textbf{4{,}234} & 84.11 & 1{,}187 \\
   & \toolExpGAWCite + \toolname & 0.84 & 0.61 & 15{,}687 & \textbf{4{,}191} & 73.28 & 1{,}310 \\
\cmidrule(lr){2-8}
   & \toolLimiWCite  & 0.85 & 0.65 & 38{,}049 & 13{,}338 & 64.95 & -- \\
   & \toolname & 0.85 & 0.64 & 38{,}049 & \textbf{4{,}294} & 87.84 & 1{,}901 \\
   & \toolLimiWCite  + \toolname & 0.85 & 0.65 & 13{,}338 & \textbf{4{,}515} & 66.15 & 1{,}468 \\
\midrule
  \multirow{9}{*}{AC2} & \toolThemisWCite& 0.84 & 0.60 & 7{,}532 & 7{,}891 & -4.77 & -- \\
   & \toolname & 0.85 & 0.61 & 7{,}532 & \textbf{4{,}140} & 45.03 & 1{,}309 \\
   & \toolThemisWCite + \toolname & 0.84 & 0.60 & 7{,}891 & \textbf{4{,}236} & 46.32 & 1{,}259 \\
\cmidrule(lr){2-8}
   & \toolExpGAWCite & 0.77 & 0.65 & 27{,}002 & 19{,}148 & 29.09 & -- \\
   & \toolname & 0.85 & 0.61 & 27{,}002 & \textbf{4{,}101} & 84.81 & 1{,}377 \\
   & \toolExpGAWCite + \toolname & 0.77 & 0.65 & 19{,}148 & \textbf{3{,}565} & 81.38 & 2{,}497 \\
\cmidrule(lr){2-8}
   & \toolLimiWCite  & 0.82 & 0.42 & 29{,}132 & 16{,}940 & 41.85 & -- \\
   & \toolname & 0.85 & 0.61 & 29{,}132 & \textbf{3{,}838} & 86.83 & 2{,}179 \\
   & \toolLimiWCite  + \toolname & 0.82 & 0.42 & 16{,}940 & \textbf{3{,}922} & 76.85 & 1{,}999 \\
\midrule
  \multirow{9}{*}{AC3} & \toolThemisWCite& 0.85 & 0.64 & 7{,}525 & 7{,}094 & 5.73 & -- \\
   & \toolname & 0.85 & 0.66 & 7{,}525 & \textbf{4{,}290} & 42.99 & 1{,}244 \\
   & \toolThemisWCite + \toolname & 0.85 & 0.64 & 7{,}094 & \textbf{4{,}754} & 32.99 & 988 \\
\cmidrule(lr){2-8}
   & \toolExpGAWCite & 0.84 & 0.62 & 20{,}481 & 24{,}652 & -20.37 & -- \\
   & \toolname & 0.85 & 0.66 & 20{,}481 & \textbf{4{,}235} & 79.32 & 1{,}317 \\
   & \toolExpGAWCite + \toolname & 0.84 & 0.62 & 24{,}652 & \textbf{4{,}121} & 83.28 & 1{,}495 \\
\cmidrule(lr){2-8}
   & \toolLimiWCite  & 0.85 & 0.64 & 46{,}081 & 11{,}014 & 76.10 & -- \\
   & \toolname & 0.85 & 0.66 & 46{,}081 & \textbf{4{,}233} & 90.81 & 2{,}478 \\
   & \toolLimiWCite  + \toolname & 0.85 & 0.64 & 11{,}014 & \textbf{4{,}120} & 62.59 & 1{,}749 \\
\midrule
  \multirow{9}{*}{AC4} & \toolThemisWCite& 0.84 & 0.65 & 7{,}491 & 7{,}999 & -6.78 & -- \\
   & \toolname & 0.85 & 0.62 & 7{,}491 & \textbf{3{,}957} & 47.18 & 1{,}512 \\
   & \toolThemisWCite + \toolname & 0.84 & 0.65 & 7{,}999 & \textbf{4{,}151} & 48.11 & 1{,}363 \\
\cmidrule(lr){2-8}
   & \toolExpGAWCite & 0.84 & 0.66 & 21{,}503 & 18{,}465 & 14.13 & -- \\
   & \toolname & 0.85 & 0.62 & 21{,}503 & \textbf{4{,}094} & 80.96 & 1{,}420 \\
   & \toolExpGAWCite + \toolname & 0.84 & 0.66 & 18{,}465 & \textbf{4{,}601} & 75.08 & 1{,}096 \\
\cmidrule(lr){2-8}
   & \toolLimiWCite  & 0.83 & 0.67 & 38{,}043 & 34{,}463 & 9.41 & -- \\
   & \toolname & 0.85 & 0.62 & 38{,}043 & \textbf{3{,}683} & 90.32 & 2{,}480 \\
   & \toolLimiWCite  + \toolname & 0.83 & 0.67 & 34{,}463 & \textbf{4{,}142} & 87.98 & 3{,}008 \\
\midrule
  \multirow{9}{*}{AC5} & \toolThemisWCite& 0.84 & 0.65 & 7{,}206 & 7{,}202 & 0.06 & -- \\
   & \toolname & 0.85 & 0.66 & 7{,}206 & \textbf{4{,}054} & 43.74 & 1{,}358 \\
   & \toolThemisWCite + \toolname & 0.84 & 0.65 & 7{,}202 & \textbf{4{,}331} & 39.86 & 1{,}229 \\
\cmidrule(lr){2-8}
   & \toolExpGAWCite & 0.83 & 0.62 & 21{,}798 & 22{,}543 & -3.42 & -- \\
   & \toolname & 0.85 & 0.66 & 21{,}798 & \textbf{4{,}293} & 80.31 & 1{,}140 \\
   & \toolExpGAWCite + \toolname & 0.83 & 0.62 & 22{,}543 & \textbf{4{,}320} & 80.84 & 1{,}518 \\
\cmidrule(lr){2-8}
   & \toolLimiWCite  & 0.82 & 0.66 & 41{,}114 & 21{,}810 & 46.95 & -- \\
   & \toolname & 0.85 & 0.66 & 41{,}114 & \textbf{4{,}066} & 90.11 & 2{,}433 \\
   & \toolLimiWCite  + \toolname & 0.82 & 0.66 & 21{,}810 & \textbf{4{,}244} & 80.54 & 2{,}351 \\
\midrule
  \multirow{9}{*}{AC6} & \toolThemisWCite& 0.84 & 0.67 & 7{,}559 & 10{,}522 & -39.20 & -- \\
   & \toolname & 0.85 & 0.61 & 7{,}559 & \textbf{3{,}851} & 49.05 & 1{,}666 \\
   & \toolThemisWCite + \toolname & 0.84 & 0.67 & 10{,}522 & \textbf{3{,}042} & 71.09 & 2{,}495 \\
\cmidrule(lr){2-8}
   & \toolExpGAWCite & 0.85 & 0.63 & 25{,}959 & 9{,}502 & 63.40 & -- \\
   & \toolname & 0.85 & 0.61 & 25{,}959 & \textbf{3{,}803} & 85.35 & 1{,}665 \\
   & \toolExpGAWCite + \toolname & 0.85 & 0.63 & 9{,}502 & \textbf{4{,}169} & 56.13 & 1{,}249 \\
\cmidrule(lr){2-8}
   & \toolLimiWCite  & 0.85 & 0.67 & 34{,}822 & 15{,}302 & 56.06 & -- \\
   & \toolname & 0.85 & 0.61 & 34{,}822 & \textbf{3{,}854} & 88.93 & 2{,}206 \\
   & \toolLimiWCite  + \toolname & 0.85 & 0.67 & 15{,}302 & \textbf{4{,}338} & 71.65 & 1{,}854 \\
\midrule
  \multirow{3}{*}{BM1} & \toolThemisWCite& 0.88 & 0.00 & 3{,}136 & 3{,}788 & -20.79 & -- \\
   & \toolname & 0.89 & 0.48 & 3{,}136 & \textbf{1{,}076} & 65.69 & 479 \\
   & \toolThemisWCite + \toolname & 0.88 & 0.00 & 3{,}788 & \textbf{1{,}161} & 69.35 & 499 \\
\midrule
  \multirow{3}{*}{BM2} & \toolThemisWCite& 0.88 & 0.00 & 3{,}291 & 3{,}638 & -10.54 & -- \\
   & \toolname & 0.89 & 0.54 & 3{,}291 & \textbf{1{,}063} & 67.70 & 489 \\
   & \toolThemisWCite + \toolname & 0.88 & 0.00 & 3{,}638 & \textbf{1{,}073} & 70.51 & 512 \\
\midrule
  \multirow{3}{*}{BM3} & \toolThemisWCite& 0.88 & 0.00 & 3{,}506 & 3{,}573 & -1.91 & -- \\
   & \toolname & 0.88 & 0.58 & 3{,}506 & \textbf{1{,}082} & 69.14 & 513 \\
   & \toolThemisWCite + \toolname & 0.88 & 0.00 & 3{,}573 & \textbf{1{,}076} & 69.89 & 513 \\
\midrule
  \multirow{3}{*}{BM4} & \toolThemisWCite& 0.23 & 0.23 & 3{,}121 & 3{,}445 & -10.38 & -- \\
   & \toolname & 0.90 & 0.53 & 3{,}121 & \textbf{1{,}060} & 66.04 & 470 \\
   & \toolThemisWCite + \toolname & 0.23 & 0.23 & 3{,}445 & \textbf{1{,}034} & 69.99 & 486 \\
\bottomrule
\end{tabular}%
&
\begin{tabular}[t]{l l | r r | r r r r}
\toprule
Model & Mitigation & Acc & F1 & $\#\text{IDI}_\text{bef}$ & $\#\text{IDI}_\text{aft}$ & Red.\ (\%) & $\#FP_{IDI}$ \\
\midrule
  \multirow{9}{*}{AC7} & \toolThemisWCite& 0.84 & 0.68 & 7{,}445 & 8{,}140 & -9.34 & -- \\
   & \toolname & 0.85 & 0.64 & 7{,}445 & \textbf{4{,}058} & 45.49 & 1{,}437 \\
   & \toolThemisWCite + \toolname & 0.84 & 0.68 & 8{,}140 & \textbf{3{,}996} & 50.91 & 1{,}826 \\
\cmidrule(lr){2-8}
   & \toolExpGAWCite & 0.76 & 0.00 & 23{,}326 & 12{,}311 & 47.22 & -- \\
   & \toolname & 0.85 & 0.64 & 23{,}326 & \textbf{4{,}105} & 82.40 & 1{,}425 \\
   & \toolExpGAWCite + \toolname & 0.76 & 0.00 & 12{,}311 & \textbf{3{,}369} & 72.63 & 1{,}842 \\
\cmidrule(lr){2-8}
   & \toolLimiWCite  & 0.76 & 0.00 & 37{,}705 & 11{,}208 & 70.27 & -- \\
   & \toolname & 0.85 & 0.64 & 37{,}705 & \textbf{4{,}001} & 89.39 & 2{,}419 \\
   & \toolLimiWCite  + \toolname & 0.76 & 0.00 & 11{,}208 & \textbf{3{,}094} & 72.39 & 2{,}112 \\
\midrule
  \multirow{9}{*}{AC8} & \toolThemisWCite& 0.84 & 0.56 & 7{,}566 & 7{,}689 & -1.63 & -- \\
   & \toolname & 0.83 & 0.66 & 7{,}566 & \textbf{4{,}302} & 43.14 & 1{,}277 \\
   & \toolThemisWCite + \toolname & 0.84 & 0.56 & 7{,}689 & \textbf{3{,}927} & 48.93 & 1{,}645 \\
\cmidrule(lr){2-8}
   & \toolExpGAWCite & 0.84 & 0.64 & 20{,}053 & 11{,}306 & 43.62 & -- \\
   & \toolname & 0.83 & 0.66 & 20{,}053 & \textbf{4{,}411} & 78.00 & 1{,}093 \\
   & \toolExpGAWCite + \toolname & 0.84 & 0.64 & 11{,}306 & \textbf{4{,}075} & 63.96 & 1{,}451 \\
\cmidrule(lr){2-8}
   & \toolLimiWCite  & 0.85 & 0.67 & 24{,}386 & 17{,}034 & 30.15 & -- \\
   & \toolname & 0.83 & 0.66 & 24{,}386 & \textbf{4{,}188} & 82.83 & 2{,}346 \\
   & \toolLimiWCite  + \toolname & 0.85 & 0.67 & 17{,}034 & \textbf{4{,}362} & 74.39 & 1{,}733 \\
\midrule
  \multirow{9}{*}{AC9} & \toolThemisWCite& 0.84 & 0.68 & 8{,}348 & 8{,}260 & 1.05 & -- \\
   & \toolname & 0.83 & 0.56 & 8{,}348 & \textbf{3{,}642} & 56.37 & 1{,}927 \\
   & \toolThemisWCite + \toolname & 0.84 & 0.68 & 8{,}260 & \textbf{3{,}691} & 55.31 & 1{,}980 \\
\cmidrule(lr){2-8}
   & \toolExpGAWCite & 0.84 & 0.65 & 20{,}006 & 8{,}120 & 59.41 & -- \\
   & \toolname & 0.83 & 0.56 & 20{,}006 & \textbf{3{,}776} & 81.13 & 1{,}727 \\
   & \toolExpGAWCite + \toolname & 0.84 & 0.65 & 8{,}120 & \textbf{4{,}391} & 45.92 & 1{,}168 \\
\cmidrule(lr){2-8}
   & \toolLimiWCite  & 0.83 & 0.58 & 51{,}399 & 22{,}330 & 56.56 & -- \\
   & \toolname & 0.83 & 0.56 & 51{,}399 & \textbf{3{,}696} & 92.81 & 2{,}379 \\
   & \toolLimiWCite  + \toolname & 0.83 & 0.58 & 22{,}330 & \textbf{4{,}160} & 81.37 & 2{,}054 \\
\midrule
  \multirow{9}{*}{AC10} & \toolThemisWCite & 0.85 & 0.63 & 7{,}509 & 7{,}250 & 3.45 & -- \\
   & \toolname & 0.78 & 0.33 & 7{,}509 & \textbf{4{,}357} & 41.98 & 1{,}164 \\
   & \toolThemisWCite + \toolname & 0.85 & 0.63 & 7{,}250 & \textbf{4{,}129} & 43.05 & 1{,}339 \\
\cmidrule(lr){2-8}
   & \toolExpGAWCite & 0.84 & 0.57 & 28{,}342 & 27{,}314 & 3.63 & -- \\
   & \toolname & 0.85 & 0.66 & 28{,}342 & \textbf{4{,}220} & 85.11 & 1{,}218 \\
   & \toolExpGAWCite + \toolname & 0.84 & 0.57 & 27{,}314 & \textbf{3{,}610} & 86.78 & 1{,}803 \\
\cmidrule(lr){2-8}
   & \toolLimiWCite  & 0.83 & 0.65 & 42{,}942 & 21{,}551 & 49.81 & -- \\
   & \toolname & 0.85 & 0.66 & 42{,}942 & \textbf{4{,}141} & 90.36 & 2{,}175 \\
   & \toolLimiWCite  + \toolname & 0.83 & 0.65 & 21{,}551 & \textbf{4{,}512} & 79.06 & 1{,}932 \\
\midrule
  \multirow{9}{*}{AC11} & \toolThemisWCite& 0.84 & 0.62 & 7{,}808 & 7{,}481 & 4.19 & -- \\
   & \toolname & 0.81 & 0.67 & 7{,}808 & \textbf{4{,}529} & 42.00 & 1{,}171 \\
   & \toolThemisWCite + \toolname & 0.84 & 0.62 & 7{,}481 & \textbf{3{,}957} & 47.11 & 1{,}596 \\
\cmidrule(lr){2-8}
   & \toolExpGAWCite & 0.80 & 0.63 & 16{,}445 & 24{,}718 & -50.31 & -- \\
   & \toolname & 0.81 & 0.67 & 16{,}445 & \textbf{4{,}746} & 71.14 & 931 \\
   & \toolExpGAWCite + \toolname & 0.80 & 0.63 & 24{,}718 & \textbf{3{,}739} & 84.87 & 1{,}844 \\
\cmidrule(lr){2-8}
   & \toolLimiWCite  & 0.76 & 0.00 & 65{,}598 & 11{,}209 & 82.91 & -- \\
   & \toolname & 0.81 & 0.67 & 65{,}598 & \textbf{3{,}677} & 94.39 & 4{,}723 \\
   & \toolLimiWCite  + \toolname & 0.76 & 0.00 & 11{,}209 & \textbf{3{,}264} & 70.88 & 1{,}968 \\
\midrule
  \multirow{9}{*}{AC12} & \toolThemisWCite & 0.76 & 0.00 & 7{,}667 & 11{,}208 & -46.18 & -- \\
   & \toolname & 0.84 & 0.65 & 7{,}667 & \textbf{3{,}972} & 48.19 & 1{,}580 \\
   & \toolThemisWCite + \toolname & 0.76 & 0.00 & 11{,}208 & \textbf{3{,}092} & 72.41 & 2{,}115 \\
\cmidrule(lr){2-8}
   & \toolExpGAWCite & 0.83 & 0.61 & 23{,}401 & 18{,}645 & 20.32 & -- \\
   & \toolname & 0.84 & 0.65 & 23{,}401 & \textbf{4{,}209} & 82.01 & 1{,}432 \\
   & \toolExpGAWCite + \toolname & 0.83 & 0.61 & 18{,}645 & \textbf{3{,}820} & 79.51 & 1{,}642 \\
\cmidrule(lr){2-8}
   & \toolLimiWCite  & 0.76 & 0.00 & 74{,}169 & 11{,}208 & 84.89 & -- \\
   & \toolname & 0.84 & 0.65 & 74{,}169 & \textbf{3{,}927} & 94.71 & 2{,}898 \\
   & \toolLimiWCite  + \toolname & 0.76 & 0.00 & 11{,}208 & \textbf{3{,}086} & 72.47 & 2{,}120 \\
\midrule
  \multirow{3}{*}{BM5} & \toolThemisWCite & 0.15 & 0.21 & 3{,}288 & 3{,}862 & -17.46 & -- \\
   & \toolname & 0.89 & 0.58 & 3{,}288 & \textbf{1{,}028} & 68.73 & 582 \\
   & \toolThemisWCite + \toolname & 0.15 & 0.21 & 3{,}862 & \textbf{908} & 76.49 & 578 \\
\midrule
  \multirow{3}{*}{BM6} & \toolThemisWCite & 0.87 & 0.01 & 3{,}175 & 3{,}862 & -21.64 & -- \\
   & \toolname & 0.89 & 0.53 & 3{,}175 & \textbf{1{,}232} & 61.20 & 361 \\
   & \toolThemisWCite + \toolname & 0.87 & 0.01 & 3{,}862 & \textbf{1{,}032} & 73.28 & 572 \\
\midrule
  \multirow{3}{*}{BM7} & \toolThemisWCite & 0.57 & 0.27 & 3{,}244 & 3{,}692 & -13.81 & -- \\
   & \toolname & 0.89 & 0.56 & 3{,}244 & \textbf{1{,}074} & 66.89 & 471 \\
   & \toolThemisWCite + \toolname & 0.57 & 0.27 & 3{,}692 & \textbf{1{,}058} & 71.34 & 516 \\
\midrule
  \multirow{3}{*}{BM8} & \toolThemisWCite & 0.81 & 0.09 & 3{,}244 & 3{,}859 & -18.96 & -- \\
   & \toolname & 0.89 & 0.43 & 3{,}244 & \textbf{1{,}074} & 66.89 & 471 \\
   & \toolThemisWCite + \toolname & 0.81 & 0.09 & 3{,}859 & \textbf{1{,}006} & 73.93 & 528 \\
\bottomrule
\end{tabular}%

\end{tabular}%
}
}
\footnotesize
\end{revisionblock}
\end{table*}

\revision{
\noindent \textbf{Mitigation Results (Qualitative Differences between \toolname vs. Baselines):}
We compare \toolname against three baselines: \toolThemis~\cite{angell2018themis}, 
\toolnameExpGA~\cite{fanExpGAICSE22}, and \toolnameLimi~\cite{LIMI-XiaoISSTA2023-10.1145/3597926.3598099}.
While \toolnameExpGA and \toolnameLimi~are IDI-generation tools focused on discovering 
individual discrimination instances, \toolname targets explanation and mitigation.
\toolname operates as a post-hoc wrapper that preserves original model utility, whereas 
counterfactual retraining modifies the model itself and risks degrading accuracy or 
collapsing when discovered IDIs are out of distribution.
Crucially, \toolname is agnostic to the upstream IDI-finding method: it takes IDIs 
discovered by \toolThemis, \toolnameExpGA, or \toolnameLimi, aligns them into relational 
datasets, infers interpretable discriminatory regions, and deploys those regions as guardrails.

Table~\ref{tab:rq5_dnn_compact} reports standard classification utility metrics 
(Accuracy, F1) and four fairness-specific columns: $\#IDI_{bef}$, the number of IDIs 
before mitigation; $\#IDI_{aft}$, the number of IDIs after mitigation; $Red~(\%)$, the 
percentage reduction between the two; and $\#FP_{IDI}$, the number of 
falsely rejected instances by the guardrail.
}

\begin{itemize}
    \item \revision{\textit{Data Augmentations vs. \toolname}. Across all AC models, \toolname consistently reduces unfair outputs. \toolname reduces IDIs by 42–56\% on the Adult Census models for \toolThemis  IDIs, by 71–85\% for \toolnameExpGA-based IDIs, and by 87–95\% for \toolnameLimi-based IDIs.
    The larger reductions for \toolnameExpGA and \toolnameLimi~stem from the greater volume (often 2-10x more vs \toolThemis) of initial IDIs these tools discover ( 16,000–74,000 vs. \toolThemis's 7,000–8,000), which provides richer relational training data helping to create more precise rules.
    On the BM models, reductions are around 61–69\%.  For BM1, we get around 65.7\% reduction from 3,136 to 1,076. For BM3, the reduction is from 3506 to 1082, or around 69.1\%. 
    }

    \item \textit{$\#FP_{IDI}$ (false-positive guardrails)}. 
     \toolname incurs the cost of selectively denying some benign inputs. For AC models, $\#FP_{IDI}$ typically ranges from \revision{ 1,100–3,000} (e.g., 1,323 for AC1; 1,927 for AC9), but each denial is offset by 2.2–2.6$\times$ as many unfair cases prevented. For instance, AC1 prevents 3,062 unfair outputs against 1,323 benign blocks.
     %

    \item \revision{\textit{Data Augmentations with Guardrails}. 
    %
    When \toolname applied after retraining with the baseline techniques (\textit{tool} $+$ \toolname), reliably lowers $\#IDI$  well below the unguarded retrained (\textit{tool}) models by 39-84\% (with \toolThemis AC3: reduced from 7094 to 4754, for AC6 reduced from 10522 to 3042, for AC7 reduced from 8140 to 3996). Also, when CF retraining reduce IDIs e.g., AC1 with \toolnameExpGA  reduces from 26,647 to 15,687), adding \toolname guardrails further reduces IDIs to 4,191 (73\% beyond retrained \toolnameExpGA alone). 
    }

\end{itemize}

\begin{answerbox}
\textbf{Answer RQ5:} 
\revision{
\toolname generalizes to improve individual fairness for black-box deep neural networks. 
The full pipeline of data curation, rule extraction, and guardrail deployment produces interpretable fairness invariants with high confidence (0.83–0.88) and substantial discrimination coverage (41–84\%) on DNNs.
\toolname consistently reduces individual discriminatory instances by at least 40\% and up to 84\%, significantly outperforming the baseline mitigation techniques.
}

\end{answerbox}

\section{Discussions}
\label{sec:discuss}
\revision{
\noindent \textit{Runtime and Performance Overhead.}
For symbolic programs, the full \toolname pipeline (IDI finding, data curation, 
interpretable model training, rule extraction, and guardrail application) completes 
in under 120 seconds per program.
For DNN benchmarks, Stage 1 IDI finding is the primary overhead, capped at 60 minutes 
(3,600 seconds) per model for all evaluated tools.
The remaining pipeline stages are completed in under 30 seconds per model.

\vspace{0.25 em}
\noindent \textit{Component-Wise Ablation Analysis.}
Although \toolname's evaluation does not use a single ablation table, each RQ is 
structured to isolate the contribution of one pipeline component while holding others fixed.
For example, RQ1 ablates \textit{data curation}: fixing the learner (CART) and varying the alignment 
strategy (DCNE, DCVE, DCHE vs.\ AFT and most-frequent baselines) to measure the 
contribution of relational alignment to localization quality.
Similarly, RQ2 ablates the \textit{interpretable learner}: fixing the curation method and 
comparing tree-based algorithms to isolate the impact of learner choice.
}

\vspace{0.25 em}
\noindent \textit{Limitation}. 
\revision{
\toolname can be integrated into existing ML pipelines as a post-processing wrapper 
requiring no modification to the underlying model.
After the DPuT produces a prediction, the guardrail module checks whether the input 
falls within a known discriminatory region and, if so, defers the decision to human 
review, making \toolname compatible with any classifier.
However, several limitations bound the current scope of applicability. First, \toolname should be viewed as an auditing and guardrail-synthesis tool rather 
than a proof of global fairness: its rules describe unfair regions represented in the 
sampled relational dataset, and repeated audits may be needed as the DPuT or data 
distribution evolves.
Second, rule quality is bounded by the coverage of the IDI-finding stage; input 
regions not represented during training---whether fair or discriminatory---will 
not be captured by the extracted rules.
If the IDI-finding stage discovers no IDIs, \toolname cannot proceed, though this 
outcome does not itself guarantee model fairness.
Third, for natural language or vision systems, protected-attribute changes can affect 
semantics in subtle ways; applying \toolname in such settings would require 
domain-specific counterfactual generators and semantic-validity filters.
We therefore do not claim that the current implementation directly addresses fairness 
debugging for all large-scale systems. 
}

\vspace{0.25 em}
\noindent \textit{Threats to Validity}. To ensure reproducibility and deterministic results, we repeat experiments multiple times, reporting averages and standard deviations. To mitigate the impact of randomness in fairness testing and decision tree initialization, we employ seed values, acknowledging the resulting seed dependency. Regarding generalizability, we evaluate our framework on 20 DNNs and diverse symbolic programs, including logic-based and scoring-based models. Internal validity concerns related to hyperparameter selection for IDI searching and interpretable model training are addressed by following established standards. Finally, to prevent variability from sampling symbolic datasets, we generate and fix our datasets once to maintain consistency across all experimental runs.
%

\revision{
\vspace{0.25 em}
\noindent \textit{Validity of Counterfactual Inputs.}
Counterfactual generation in \toolname follows a two-step process to ensure semantic validity.
First, the protected attribute is flipped within its valid domain (e.g., sex, ethnicity).
Second, any non-protected attributes causally dependent on the protected attribute are updated according to domain-specific symbolic rules.
For symbolic DPuTs, these constraints are derived directly from the benchmark program logic. 
For DNN benchmarks, we inherit input domains and preprocessing constraints from the corresponding dataset (e.g., flipping sex from male to female triggers an update of the relationship attribute).
However, for DNN benchmarks, a potential limitation is that causal dependencies between features may not be fully known, meaning some generated counterfactuals may be invalid due to the limitations of preprocessing rules. 

\vspace{0.25 em}
\noindent \textit{Validity of Explanations.:}
\toolname Explanations are quantitatively validated in different parts of the paper. 
In RQ3 and RQ5, we measure impurity, confidence, and coverage of the extracted rules against IDI localization and coverage metrics. 
Qualitative correctness is confirmed by benchmarking extracted invariants against ground-truth symbolic logic.
In RQ4–RQ5, we demonstrate that applying these rules as guardrails effectively reduces IDIs, which would not be possible if the explanations were incorrect.
A limitation is that for DNNs, where the ground-truth discriminatory logic is unknown, we can only validate explanations indirectly through their mitigation effectiveness.
}

\vspace{0.25 em}
\revision{
\noindent \textit{Interpretability in Symbolic Setting.}
Interpretability is not for identifying protected attributes, but for mapping discriminatory logic to regions of non-protected features. 
This enables generalization from point-wise violations to systematic patterns, which is critical for deploying automated mitigation guardrails, especially in a black-box setting.
Besides, interpretability in a symbolic setting serves as a validation step to assess whether our approach accurately explains the ground-truth discrimination. 
}

\revision{
\vspace{0.25 em}
\noindent \textit{Concrete Examples of Fairness Violations, Rules, and Mitigation.}
To illustrate \toolname in a realistic fairness-debugging workflow, we consider a 
COMPAS-like recidivism risk assessment tool, which has been widely documented to 
produce racially and sexually disparate predictions~\cite{compas-article}.
Our symbolic benchmark PG2 encodes a simplified recidivism scoring rule 
(PG2 Listing in \cite{Akash-REMI-ISSTA-2026-SupplMat} from ~\cite{rudin2019stop}). 

\begin{itemize}
    \item \noindent\textit{Stage 1: IDI Discovery.} From 6,908 individuals sampled from the preprocessed COMPAS  dataset~\cite{compas-dataset, ZhongNIPS2023GSGAMS-10.5555/3666122.3668598, larson2016we},  
    \toolname find 206 IDIs by flipping \texttt{sex} (male $\leftrightarrow$ female).

    \item \noindent\textit{Stage 2: Relational Data Curation.} Using DCNE, \toolname constructs a relational dataset of 6,908 instances, each labeled fair ($+$) or discriminatory ($-$) based on relational data points.
    \item \noindent\textit{Stage 3: Rule Extraction.} A CART decision tree trained on the relational labels extracts the rule  \texttt{\{age $\leq$ 20.5 $\land$ priors\_count $\leq$ 3.5\}},  which captures all 206 IDIs with high confidence.
    \item \noindent\textit{Stage 4: Guardrail Mitigation.} Deploying this rule as a guardrail eliminates all 206 IDIs, achieving a full reduction in discriminatory behavior without changing the logic of the decision-making program.    
\end{itemize}
}

\section{Related Work}
\label{sec:related}
\noindent\textbf{Fairness Testing:} 
When Galhotra et al.~\cite{GalhotraThemisFSE2017} proposed and popularized the causal fairness definition for individual fairness testing, a substantial line of work ~\cite{angell2018themis,agarwal2018automated,SG-AggarwalESEC/FSE201910.1145/3338906.3338937,fanExpGAICSE22,10.1109/ICSE55347.2025.00070} focused on discovering discriminatory inputs (IDIs) through testing at scale following that fairness definition.
Some of the tools e.g., \textsc{Themis}~\cite{angell2018themis}, \textsc{AEQUITAS}~\cite{udeshi2018automated}, 
\textsc{SG}~\cite{SG-AggarwalESEC/FSE201910.1145/3338906.3338937},
\textsc{ExpGA}~\cite{fanExpGAICSE22}, 
\textsc{LIMI}~\cite{LIMI-XiaoISSTA2023-10.1145/3597926.3598099},
\textsc{AFT}~\cite{AFT-zhao-ase2024}, 
considered black-box settings, where the inner working knowledge of classifiers was not necessary for finding ID instances.
Some other test generation algorithms 
\textsc{ADF}~\cite{zhang2020white}, 
\textsc{EIDIG}~\cite{EDIG-Zhang-ISSTA2021-10.1145/3460319.3464820},
\textsc{NeuronFair}~\cite{ZhengICSE2022NeuronFair},
\textsc{DICE}~\cite{Monjezi2023InformationTheoreticTA}, 
\textsc{MAFT}~\cite{MAFT-Wang-ICSE2024-10.1145/3597503.3639181}
considers a white-box setting, which requires knowledge of the classifier being tested
Many of these tools can be used in the first stage of our framework to find IDIs for DPuTs under test if the DPuT matches the target model for these tools.
\noindent The problem of testing for group fairness has been explored extensively in existing literature \cite{bellamy2019ai, chakraborty2020fairway, DBLP:conf/icse/ZhangH21, 10.1145/3468264.3468537, tizpaz2022fairness-10.1145/3510003.3510202, 10.1145/3540250.3549093}.

\noindent \textbf{Formal Methods.} Formal tools and techniques have been significantly studied in the literature~\cite{10.1109/ICSE55347.2025.00016,11334575,10.1007/978-3-031-37703-7_16,kim2026analyzingfairnessneuralnetwork,Monjeze26ICSE}
The \textsc{FairSquare} tool \cite{fairsquare17oopsla} targets probabilistic programs, employing volume-based computations to verify their fairness properties.
A formal methodology for certifying individual fairness in standard machine learning architectures was established by John et al. \cite{john2020verifying}.
\textsc{Fairify} \cite{BiswasICSE2023Fairify-10.1109/ICSE48619.2023.00134} examines individual fairness and its variants in neural networks by translating pre-trained models into Satisfiability Modulo Theories (SMT) problems.

\noindent\textbf{Explainable-AI and Interpretability:}
Riberio et al.\cite{ribeiro2016should} provide a local explanation model, LIME, to explain a locally faithful non-linear model via a sparse linear model
Lundberg. et al., \cite{LundbergShaplyNIPS17} propose a unified approach with SHAP and SHAPly values to interpret the global behavior of the model.
Mothilal et al. \cite{MothilalFAT20DiCE} provide a tool called DiCE, which utilizes diverse counterfactual explanations to understand the decision boundary of the model, more specifically, changes in which feature will lead to the flip of the decision from one side of the boundary to the other side.
\textsc{FairLay-ML}~\cite{10.1109/ICSE-Companion66252.2025.00016} is a debugging tool to explain the fairness implications of data-driven software.
\textsc{Parfait-ML}~\cite{tizpaz2022fairness-10.1145/3510003.3510202,10.1145/3663533.3664040} explores the explanation behind hyperparameter configuration, which might lead to unfair models via decision trees.
Instead, we used interpretable models to explain the root cause of unfairness.


\noindent \textbf{Bias Mitigation} 
To address bias in machine learning outputs, several researchers have proposed different types of mitigation algorithms \cite{zhang2018mitigating,6413831,agarwal2018reductions}.
Our work is closer to post-processing techniques~\cite{dasu2024neufair,10.1145/3510003.3510080,10.1145/3617168,10.1145/3540250.3549103,10.1145/3510003.3510087,Dasu26Attn}. While these works aim to modify the decision logic of black-box models, our approach leverages guardrails, derived from the explanation models, to improve the fairness of symbolic and data-driven black-box software.

\section{Conclusion and Future Work}
\label{sec:conclusion}
In this paper, we presented \toolname, a novel framework that treats individual fairness as a relational invariant discovery problem. By transforming counterfactual pairs into a relational dataset, \toolname successfully localizes discriminatory regions using interpretable, rule-based explainers. These extracted fairness invariants provide both a precise explanation of the root cause of discrimination and a robust mitigation strategy via real-time guardrails. Our evaluation across symbolic, scoring, and deep neural network programs demonstrates that \toolname is highly effective and outperforms the state-of-the-art baselines. Future work includes extending this relational framework to unstructured domains, such as natural language~\cite{dahal-etal-2026-investigating}.

\section{Data Availability}
\label{sec:data-availability}
Our open-source tool \toolname with all experimental subjects is available at Figshare~\cite{Akash-REMI-ISSTA-2026-SupplMat} and \href{https://github.com/ranit43/REMI_Fair}{Github}.

\section*{Acknowledgments}
This project has been partially supported by NSF under grants CCF-2536640 and CNS-2230061. 

\bibliographystyle{ACM-Reference-Format}
\bibliography{references}


\end{document}